\documentclass{vldb}
\usepackage{graphicx}
\usepackage{balance}  

\vldbTitle{}
\vldbAuthors{Zhiwei et al.}
\vldbDOI{https://doi.org/10.14778/xxxxxxx.xxxxxxx}
\vldbVolume{12}
\vldbNumber{xxx}
\vldbYear{2019}

\usepackage{booktabs} 
\usepackage{xspace}
\usepackage{listings}
\usepackage{url}
\usepackage{float}
\usepackage{tabularx,colortbl}
\usepackage{ragged2e}
\usepackage{algorithm}
\usepackage{wrapfig}
\usepackage{lipsum}
\usepackage{float}
\usepackage{graphicx,graphics}
\usepackage{subfig}
\usepackage[svgnames,table]{xcolor}
\usepackage[noend]{algpseudocode}
\usepackage{thmtools, thm-restate}
\usepackage{mathrsfs}
\usepackage{tikz}
\usepackage{comment}
\usepackage{makecell}
\usepackage{multirow}

\usetikzlibrary{positioning}
\usetikzlibrary{decorations.text}
\usetikzlibrary{decorations.pathmorphing}
\usetikzlibrary{arrows,petri, topaths}
\usepackage[inline]{enumitem}
\DeclareMathOperator*{\argmin}{argmin}

\makeatletter
\def\set@curr@file#1{%
  \begingroup
    \escapechar\m@ne
    \xdef\@curr@file{\expandafter\string\csname #1\endcsname}%
  \endgroup
}
\def\quote@name#1{"\quote@@name#1\@gobble""}
\def\quote@@name#1"{#1\quote@@name}
\def\unquote@name#1{\quote@@name#1\@gobble"}
\makeatother

\newcommand*{\affmark}[1][*]{\textsuperscript{#1}}

\newcommand{\revise}[1]{{\color{black}#1}}

\newcommand{\gl}{Generalization Lvl}

\newcounter{reviewcounter}

\newcommand{\inlineitem}[1][]{%
\ifnum\enit@type=\tw@
    {\descriptionlabel{#1}}
  \hspace{\labelsep}%
\else
  \ifnum\enit@type=\z@
       \refstepcounter{\@listctr}\fi
    \quad\@itemlabel\hspace{\labelsep}%
\fi}
\makeatother

\newtheorem{example}{Example}

\newcommand{\cut}[1]{}

\newenvironment{packed_item}{
\begin{itemize}
   \setlength{\itemsep}{1pt}
   \setlength{\parskip}{0pt}
   \setlength{\parsep}{0pt}
}
{\end{itemize}}

\newenvironment{packed_enum}{
\begin{enumerate}
   \setlength{\itemsep}{1pt}
  \setlength{\parskip}{0pt}
   \setlength{\parsep}{0pt}
}
{\end{enumerate}}

\newcommand{\fcard}{\textsc{card}}
\newcommand{\fcount}{\textsc{count}}
\newcommand{\fweight}{\textsc{weight}}
\newcommand{\fcost}{\textsc{cost}}

\newcommand{\introparagraph}[1]{\vspace{1mm} \noindent \textbf{#1 }}

\newcommand{\PreserveBackslash}[1]{\let\temp=\\#1\let\\=\temp}
\newcolumntype{C}[1]{>{\PreserveBackslash\centering}p{#1}}
\newcolumntype{R}[1]{>{\PreserveBackslash\raggedleft}p{#1}}
\newcolumntype{L}[1]{>{\PreserveBackslash\raggedright}p{#1}}

\begin{document}
\setlength{\abovedisplayskip}{3pt}
\setlength{\belowdisplayskip}{3pt}
\setlength{\belowcaptionskip}{-5pt}
\setlength{\textfloatsep}{10pt}
\setlength{\floatsep}{5pt}


\title{A Comparative Exploration of ML Techniques for Tuning Query Degree of Parallelism}



%
%
%
%

\numberofauthors{1} 

\author{%
%
\alignauthor
Zhiwei Fan\affmark[1], 
Rathijit Sen\affmark[2], 
Paraschos Koutris\affmark[1], 
Aws Albarghouthi\affmark[1], 
\vspace{1ex} \\
\affaddr{\affmark[1]University of Wisconsin-Madison} 
\affaddr{\affmark[2]Microsoft Gray Systems Lab}\\ 
\email{\{zhiwei, paris, aws\}@cs.wisc.edu}
\email{\{rathijit.sen\}@microsoft.com}
}


\maketitle

\date{\today}

\maketitle

\begin{abstract}
There is a large body of recent work applying machine learning (ML) techniques to query optimization and query performance prediction in relational database management systems (RDBMSs). However, these works typically ignore the effect of \textit{intra-parallelism} -- a key component used to boost the performance of OLAP queries in practice -- on query performance prediction.
In this paper, we take a first step towards filling this gap by studying the problem of \textit{tuning the degree of parallelism (DOP) via ML techniques} in Microsoft SQL Server, a popular commercial RDBMS that allows an individual query to execute using multiple cores. 

In our study, we cast the problem of DOP tuning as a {\em regression} task, and examine how several popular ML models can help with query performance prediction in a multi-core setting. 
We explore the design space and perform an extensive experimental study comparing different models against a list of performance metrics, testing how well they generalize in different settings: $(i)$ to queries from the same template, $(ii)$ to queries from a new template, $(iii)$ to instances of different scale, and $(iv)$ to different instances and queries.  
Our experimental results show that a simple featurization of the input query plan that ignores cost model estimations can accurately predict query performance, capture the speedup trend with respect to the available parallelism, as well as help with automatically choosing an optimal per-query DOP. 

\end{abstract}

\section{Introduction}
\label{sec:intro}

The rise of cloud-computing platforms has offered new capabilities and benefits to users, including the capability to provision the appropriate amount of resources necessary for a given task. In order to effectively utilize this capability, Infrastructure-as-a-Service (IaaS) and Platform-as-a-Service (PaaS) users need tools and techniques to quantify the tradeoffs between resource costs and performance benefits. Such techniques are also invaluable to PaaS and Software-as-a-Service (SaaS) cloud providers, since they can reduce operational costs while meeting Service-Level Agreements (SLAs). 

Identifying optimal or near-optimal resource configurations is a difficult task. The challenges are three-fold: unknown workload characteristics, a large state space, and costly evaluation of what-if scenarios. Although resource utilization may be observed for repeating jobs, one may have to resort to intrusive profiling (not feasible in production environments) or A/B testing (which requires additional resources) to estimate tradeoffs with different configurations. Resource provisioning for new jobs is even more challenging. For these reasons, configuration tuning is traditionally performed by database administrators. However, this approach is tedious, requires a high level of domain expertise, and does not scale to cloud platforms that need to handle millions of databases and different service tiers. 

\begin{figure}
\centering
\includegraphics[width=0.38\textwidth]{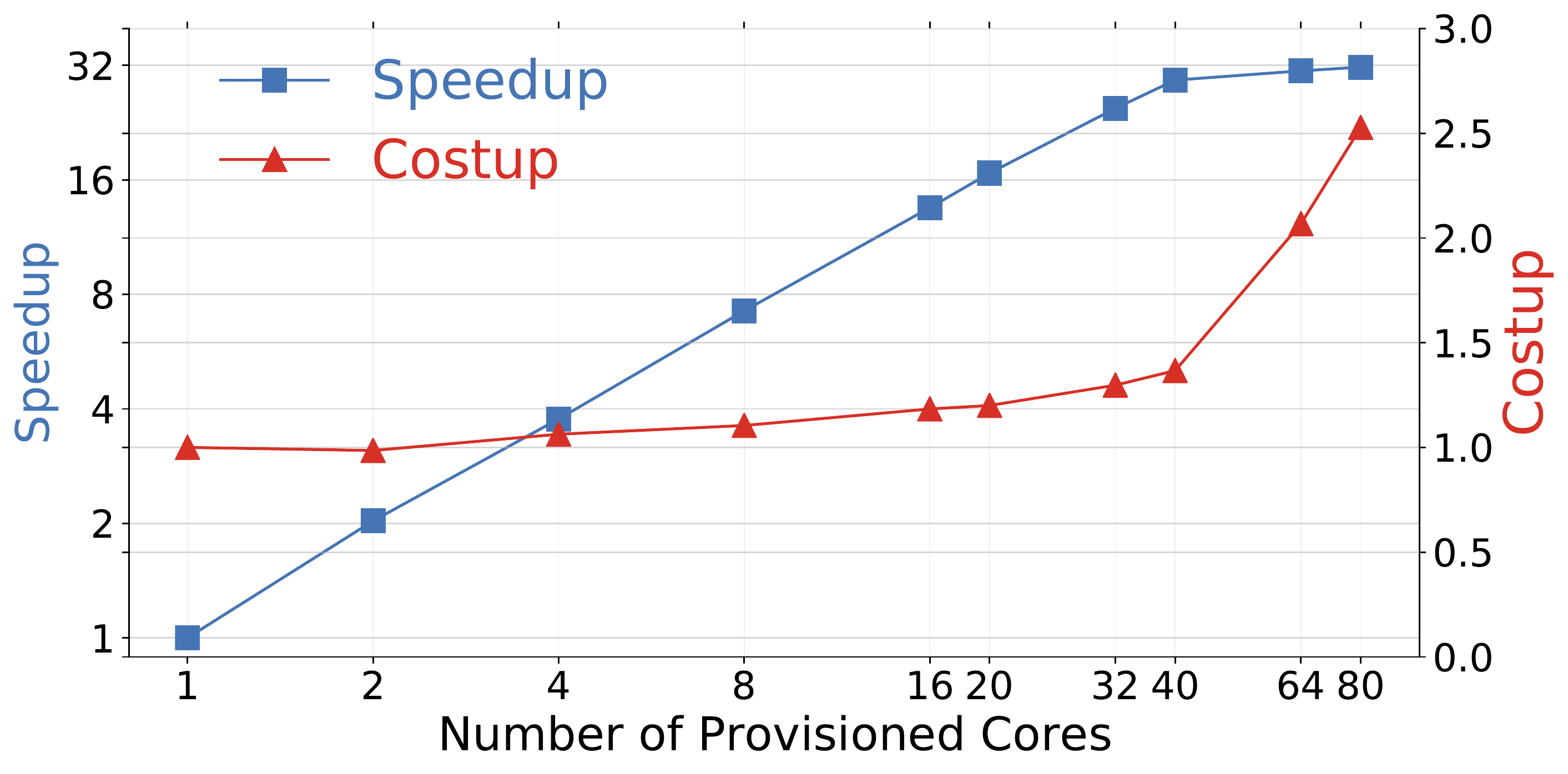}
\caption{Speedup and Costup over those with \#provisioned cores=1, DOP=1 for 22 queries in TPC-H1000.}
\label{fig:speedup-costup}
\end{figure}

In this paper, we tackle this problem by studying how {\em Machine Learning (ML) techniques can be applied to estimate cost-performance tradeoffs for query processing in an RDBMS}. 
To focus our study, we use Microsoft SQL Server as the RDBMS, which is available as cloud IaaS and PaaS offerings~\cite{azuresql} and as an on-premises solution~\cite{sqlserver-2019}. We also choose to study the configuration of a particular resource, the query {\em Degree Of Parallelism (DOP)}: this is the maximum number of hardware threads (logical cores) that can be used at any time for executing the query. Estimates of cost-benefit tradeoffs for different DOPs allow the provisioning for on-premises server configurations, IaaS compute sizes, and PaaS service tier selections---Azure SQL Database~\cite{azuresqldb-paas} offers different service tiers that allow users to choose a specific number of (logical) cores~\cite{azuresql-vcore-service-tiers, azuresql-vcore-resource-limits}, or a range of cores~\cite{azuresql-serverless-tier}, or a pre-configured size with cores bundled with other resources~\cite{azuresql-dtu-service-tiers, azuresql-dtu-resource-limits}.  
\vspace{-2ex}
\begin{example}
Figure~\ref{fig:speedup-costup} shows the speedup and costup of 22 TPC-H1000 queries (on non-partitioned data) on SQL Server, running on our system (details in Section~\ref{subsec:experimental-setup}), for different values of the total number of logical cores allowed for the workload. Here the speedup and cost are normalized with respect to the performance and cost for a single core (DOP$=$1). For this example, we define cost as the number of provisioned cores multiplied by the total time.
The figure assumes optimal per-query DOP selections, with DOP $\leq$ \#provisioned cores. 
 Doubling the number of cores from 20 to 40 improves performance by 76\% with a 14\% increase in cost, while a further doubling to 80 cores additionally increases performance by 8\% with an 85\% increase in cost. These estimates can be used to select service tiers depending on whether the business value of the additional performance outweighs the costs.
\end{example}

\introparagraph{Choosing a DOP }
Query DOP is a critical knob that affects query performance and resource utilization of multicore servers. However, choosing the \textit{optimal} or \textit{near-optimal} DOP is not easy. For well-parallelizable queries, performance improves as DOP increases, but diminishing returns are observed beyond some point depending on the database instance and query characteristics. 
For other queries, increasing DOP could be detrimental to performance due to synchronization overheads and additional resource requirements (e.g., memory), resulting in potential loss of business value, in addition to increasing cost of operations. 
In general, there is no single DOP choice that is optimal across all queries: as we will see in Section~\ref{sec:dop}, one can obtain significant improvements with a per-query DOP selection.
As a result, developing automated techniques that can make fine-grained resource allocation decisions can unlock further opportunities for performance benefits.

\vspace{2ex}
\noindent
Recent work \cite{akdere2012learning,  li2012robust, marcus2019plan} has explored the use of various ML techniques on the problem of query performance prediction (QPP). On the surface, the problem of DOP tuning at query-level can be reduced to QPP with DOP being considered as an additional feature of the model: given a query and a set of DOP values of interests, we can then select the DOP value such that the estimated query execution time of the ML model satisfies the specified resource and performance requirements. However, directly applying such techniques, that consider only a single DOP value (usually, 1), is not feasible in our setting. In particular, both \cite{li2012robust} and \cite{marcus2019plan} estimate the overall query resource/latency by aggregating/combining predicted per-operator latencies. But such information may not be readily available or easy to use with complex systems that have multithreaded query execution (due to the potential for inter-operator overlap), runtime optimizations, multiple processing modes (e.g., row/batch mode) per operator, etc.

Based on the above observation, we posit that it is necessary to consider \textit{the  query plan as a whole without using operator-level execution time information} for ML-based DOP tuning. 
An interesting outcome of our work is that approaches that use a simple plan representation are very effective in many cases when they are applied to DOP tuning while having been dismissed in previous relevant work such as QPP.


\introparagraph{Our Contribution}
In this work, we study the problem of automating fine-grained DOP tuning using ML-based techniques. To the best of our knowledge, this is the first work to study this problem in the context of an RDBMS. Our contributions can be summarized as follows:

\begin{itemize}[noitemsep,topsep=0pt]
\item  We emphasize the importance of DOP tuning and show the potential benefits of fine-grained per-query DOP tuning (Section~\ref{sec:dop}).  
\item We formulate the DOP tuning problem as a regression task and study how to best represent each query as a feature vector (Section~\ref{sec:main}). We perform a detailed experimental evaluation of different ML models using well-known decision support benchmarks (TPC-H \cite{poess2000new} and TPC-DS \cite{nambiar2006making}) consisting of queries generated from different query templates, database instances and scale factors. 
\item In order to evaluate the performance of the ML models, we use four generalization levels based on different application scenarios: $(i)$ to queries from the same template, $(ii)$ to queries from a new template, $(iii)$ to instances of different scale, and $(iv)$ to different instances and queries.  
We also use both \textit{task-agnostic} and \textit{task-specific} performance metrics to evaluate the utility of each model. Our experiments show that a simple while proper featurization along with tree-ensemble models is effective for DOP Tuning in most  cases. We also experimentally analyze the causes of systematic failures, and discuss about possible improvement as our future work (Section~\ref{sec:exp}).

\end{itemize}


\section{Degree of Parallelism}
\label{sec:dop}

In SQL Server, the query Degree Of Parallelism (DOP) is the maximum number of hardware threads or logical processors that can be used at any time to execute the query. 
\revise{Intra-query parallel execution is inspired by the Volcano~\cite{graefe1990encapsulation} operator model.}
The physical operators in a query plan can be both serial and parallel, but all parallel operators have the same DOP, which is fixed at the start of the query execution.
\revise{Being able to determine the optimal DOP in one shot at (or before) query startup is important since a suboptimal DOP cannot be fixed without terminating query execution and restarting it with a different DOP.}

\revise{SQL Server provides mechanisms to configure the DOP both at query optimization time and query startup time. In this work, we focus only on the latter option. The default DOP in the version of SQL Server we use, and for the machines used for our experiments, is 64. Thus, we choose to use the query performance at DOP 64 as the baseline in our experiments. More details regarding our experimental setup are in Section~\ref{subsec:experimental-setup}.
} 


\begin{figure}[ht]
\centering
\includegraphics[width=0.4\textwidth]{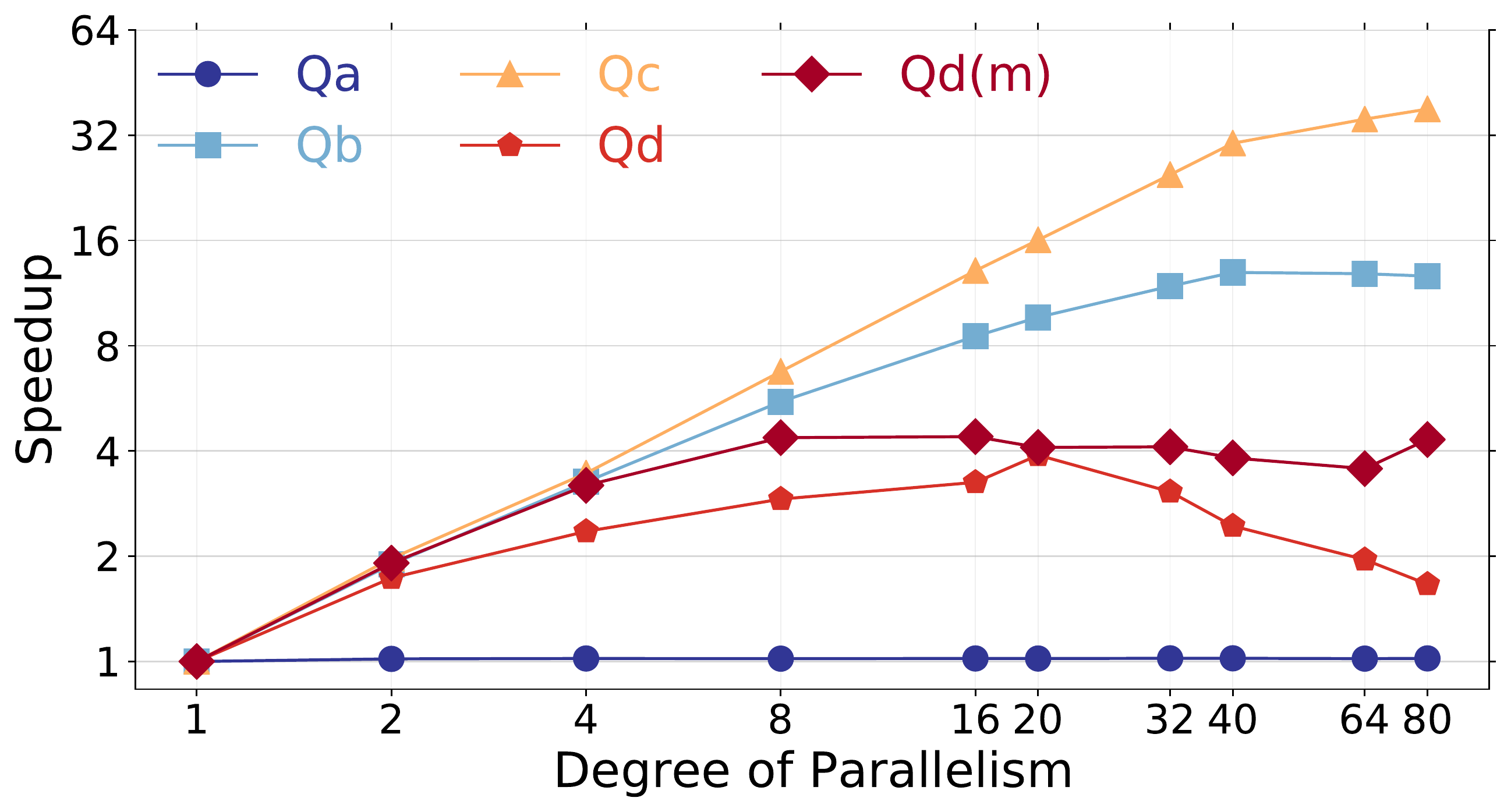}
    \caption{Performance profiles for 4 queries on DS1000}
\label{fig:generalization2-examples}
\end{figure}

As we discuss in the introduction, each query performs differently as the DOP increases. Figure~\ref{fig:generalization2-examples} shows the performance profiles for 4 different queries on TPC-DS with scale factor 1000 (see Section~\ref{sec:exp} for details about our experimental setup and workloads). Queries Qb and Qc are well-parallelizable and their performance improves as DOP increases, but with diminishing returns beyond some point. Query Qa sees no benefit from parallelism.  Query Qd loses performance at high DOPs due to disk spills caused by increased memory requirement that exceeds the default limit for a single query. Query Qd(m) shows how Qd perform differently as the DOP increases by intentionally giving enough memory to avoid disk spilling, indicating that memory is an important factor which can influence the query behavior at different DOP. One can also observe that there is no single DOP choice that is optimal across all four queries.

\begin{figure}[ht]
\centering
\includegraphics[width=0.4\textwidth]{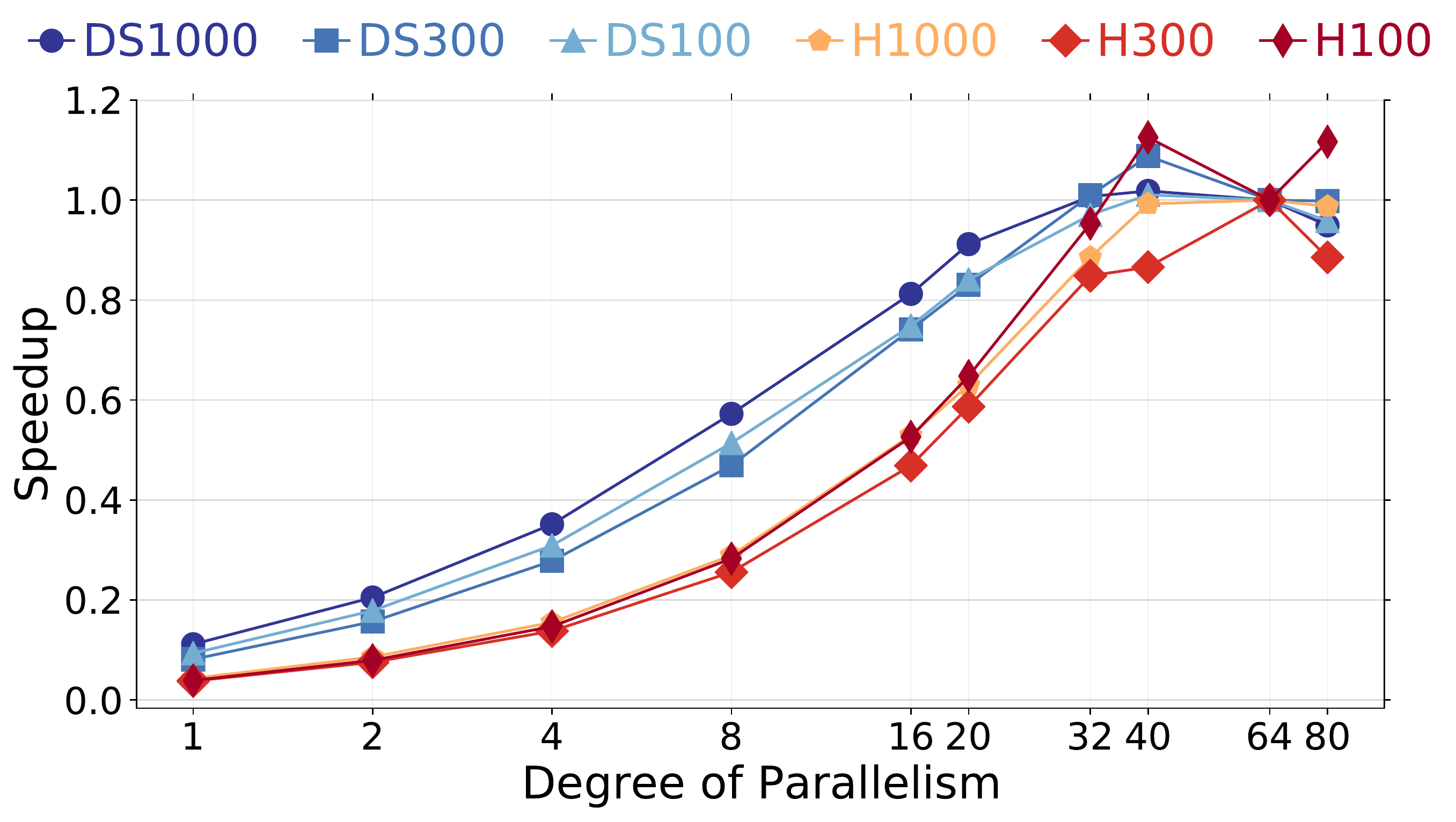}
\caption{Workload speedup over workload-level DOP=64 (default value) for different workload-level DOPs.}
\label{fig:workload-speedup}
\end{figure}

Figure~\ref{fig:workload-speedup} shows workload performance (details in Section~\ref{subsec:experimental-setup}) as a function of the DOP value, with the same DOP for all parallel queries, relative to the performance with the default DOP of 64 for our setup. For most workloads, performance improves with DOP, starting from DOP=1, and attains the peak at DOP values that are less than the default (64) and the maximum value (80). The optimal value is workload-specific and can result in substantial performance gains over the default value for some workloads.

\begin{figure}[ht]
\centering
\includegraphics[width=0.4\textwidth]{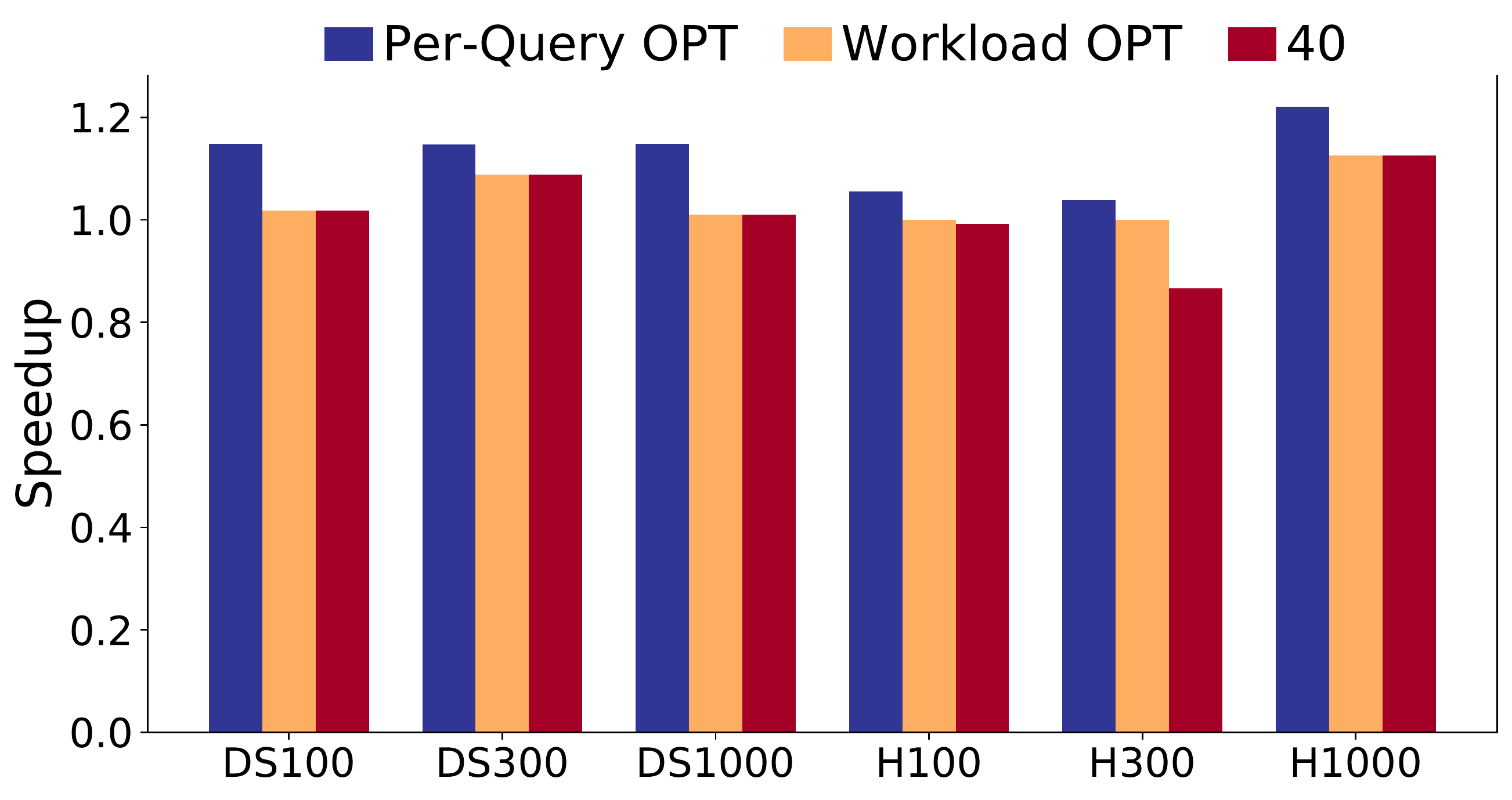}
\caption{Workload speedup (Y-Axis) over workload-level DOP=64 (default value) with per-query optimal DOPs, workload-optimal DOP, and workload-level DOP=40.}
\label{fig:workload-speedup-OPT}
\end{figure}
Further gains are possible by selecting optimal DOPs on a per-query basis. Figure~\ref{fig:workload-speedup-OPT} compares workload-level speedups possible with per-query DOP selections, per-workload DOP selections, and a static selection of DOP=40 (equal to the number of physical cores in our server). All speedups values are with respect to the static (default) value of DOP=64. In line with observations from Figure~\ref{fig:workload-speedup}, DOP=40 speeds up some workloads (e.g., H 1000), but slows down some others (e.g., H 300), and a per-workload optimal DOP can significantly improve performance (e.g., DS 300). All workloads also show additional speedups, ranging from 4\% (H 300) to 15\% (DS 1000) beyond what is possible with a workload-optimal selection.

\begin{figure}[ht]
\centering
\includegraphics[width=0.4\textwidth]{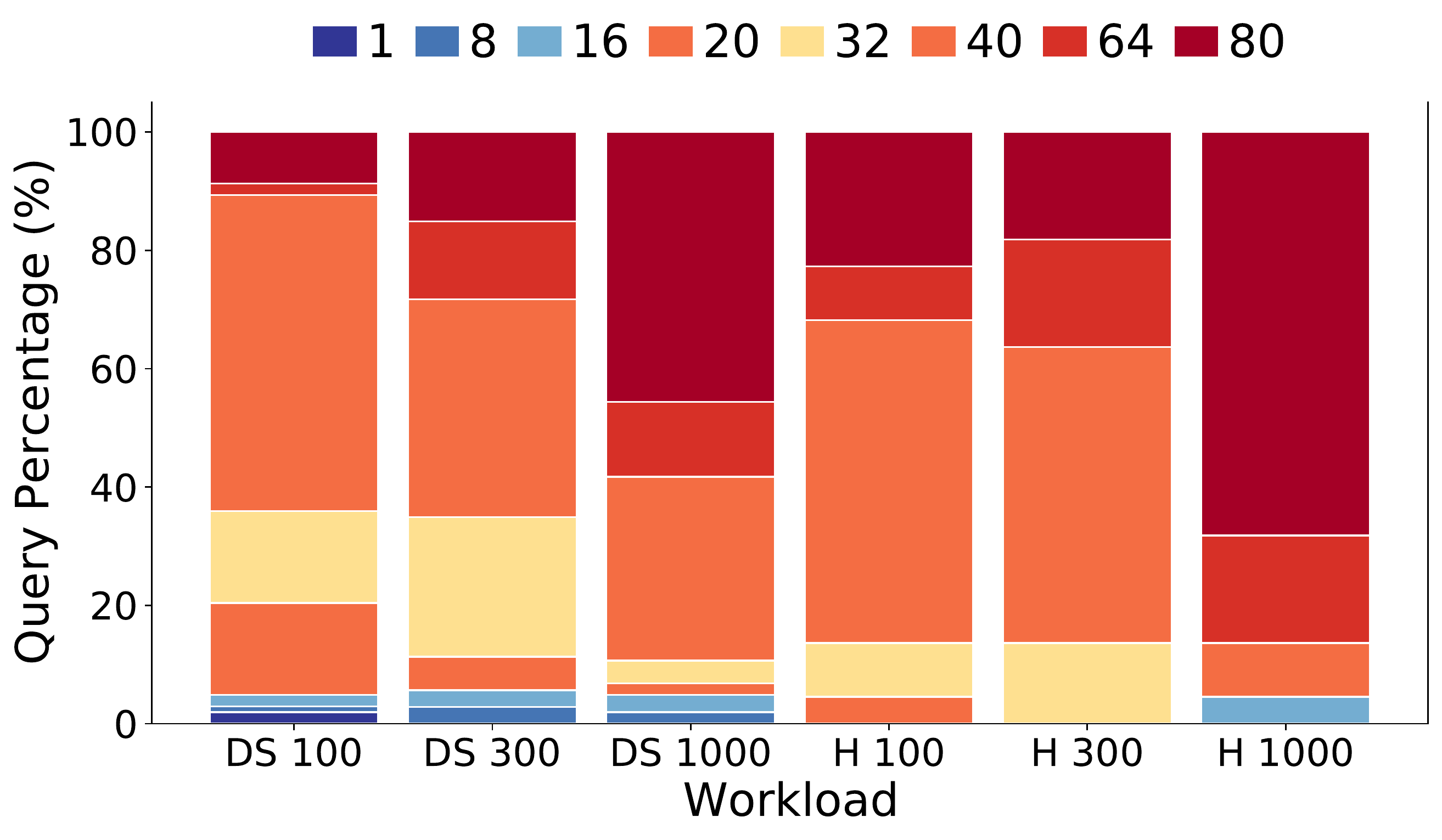}
\caption{Distribution of the percentage of queries with respect to the optimal query DOP.}
\label{fig:perquery-OPT-distrib}
\end{figure}
Just as the per-workload optimal DOP differs across workloads, the per-query optimal DOP values within each workload also show substantial variation. Figure~\ref{fig:perquery-OPT-distrib} shows more variation for the DS workloads than the H workloads, which is due to a larger variety of query templates in DS. No workload has a single per-query optimal DOP value, indicating the potential for speedup with per-query DOP selections as we observed in Figure~\ref{fig:workload-speedup-OPT}. The average and median shift towards larger DOP values as scale factors, and consequently, size of datasets increase.

\revise{
Based on the above, we can observe the following:
\begin{itemize}[noitemsep,topsep=-1pt]
\item The behaviors of different queries for different DOPs are different (Figure~\ref{fig:generalization2-examples}).
\item The optimal DOP choices for different workloads (Figure~\ref{fig:workload-speedup}) and queries are different (Figure~\ref{fig:perquery-OPT-distrib}).
\item While fine-grained DOP tuning (on a per-query level) is more challenging compared to DOP tuning at workload level, it can lead to greater performance improvement (Figure~\ref{fig:workload-speedup-OPT}). 
\end{itemize}
}

\section{The Main Components}
\label{sec:main}

In this section, we formally describe our approach to solving the DOP tuning problem. We first formally state the ML task we are solving. Then, we describe how we featurize the input data to the ML models and present the ML models we consider in this work. Finally, we discuss different ways to evaluate how well our ML models generalize under different application scenarios. 

\revise{
\introparagraph{Regression over Classification}
At first glance, it is natural to think of tuning query DOP as a classification 
problem: given a query and a set of DOP values, classify the query into the  DOP class that achieves optimal performance.
We choose instead to cast the problem as a regression problem for two reasons.
First, the performance difference of a query at neighboring DOP values can be very small  (see for example Figure~\ref{fig:generalization2-examples}), which means that accurate classification may not be possible.
Second, formulating DOP tuning as a regression problem allows us to use the learned model in applications beyond DOP selection, such as resource provisioning. We also note that the capability to predict the estimated time of a query at different DOPs might be helpful in order to perform DOP selection for concurrent query execution, which we consider as future work. 
}
\subsection{Problem Formulation}
\label{sec:task_desc}

We are given a database instance $I$ over a schema $\mathbf{R}$, and a workload $W = (P_1, P_2, \dots, P_n)$ over the instance $I$, where each $P_i$ is a compiled {\em query plan} generated by the RDBMS. For our purposes, a query plan is a tree where each node is a physical operator (e.g., Index Scan, Hash Join, Sort); additionally, for each node we are given certain information about the operator execution (e.g., estimated rows read, whether it runs in parallel mode, etc). Figure~\ref{fig:plan_example} shows an example of such a query plan. In the next section, we will show how to map the query plan $P_i$ to a feature representation $\mathsf{ftr}(P_i)$. 


For each plan $P_i$, we also have a {\em measurement} of the execution time of the plan $P_i$ over instance $I$ with DOP $d$, where $d$ comes from a set $\mathbf{D}$ of DOP values of interest. We denote this time with $t_{(P_i,d)}$. Every such measurement corresponds to a {\em training point} $(x,y)$, where $x = (\mathsf{ftr}(P_i), d)$, and $y = t_{(P_i, d)}$.

We can now cast our problem as a {\em regression} task.
In particular, the goal is to use the training points in order to learn a function $f$ (model instance) of a model $m$ that minimizes the quantity
\[ \sum^{n}_{i = 1} \sum_{d \in \mathbf{D}} \mathcal{L}(f(\mathsf{ftr}(P_i),d), t_{(P_i, d)}) \]
where $\mathcal{L}$ is a loss function. 
In this paper, we will use as a loss function 
the {\em mean squared error (MSE)},  $\mathcal{L}(a, b)= {(a-b)}^2$. We will use $\hat{t}_{(P_i,d)} = f(\mathsf{ftr}(P_i),d)$ to denote the estimated execution time using the learned model instance.


\smallskip
Using the learned model instance $f$, we can now solve the DOP tuning problem. We distinguish two subproblems. 

\introparagraph{DOP Selection at Workload Level} In practice, users in many cases configure the DOP at workload-level, in which the same DOP is selected for all the queries in the workload. In this case, the task is to choose a single DOP, 
$${d}^w_{\text{workload}} =  \argmin_{d \in \mathbf{D}} \sum_{i=1}^n{f(\mathsf{ftr}(P_i), d)}.$$


\introparagraph{DOP Selection at Per-Query Level}
Further performance gains are possible if we perform DOP selection on a per-query basis. In this case, the task is to select for each query plan $P_i$ a DOP
%
%
$${d}_i = \argmin_{d \in \mathbf{D}} f(\mathsf{ftr}(P_i), d). $$

\subsection{Featurization}
\label{sec:features}


In this section, we describe how we map a query plan $P$ to a feature vector $\mathsf{ftr}(P)$ of fixed dimension in a similar manner to what is presented in~\cite{ding2019ai}. Compared to Ding et al.~\cite{ding2019ai}, our featurization contains richer 
information by which the characteristics of a given query plan could be more accurately captured. We propose different featurization alternatives for a query plan -- we study the effects of different featurization choices in Section~\ref{sec:featurization_analysis}.

\begin{figure}[ht]
\centering
\includegraphics[width=0.5\textwidth]{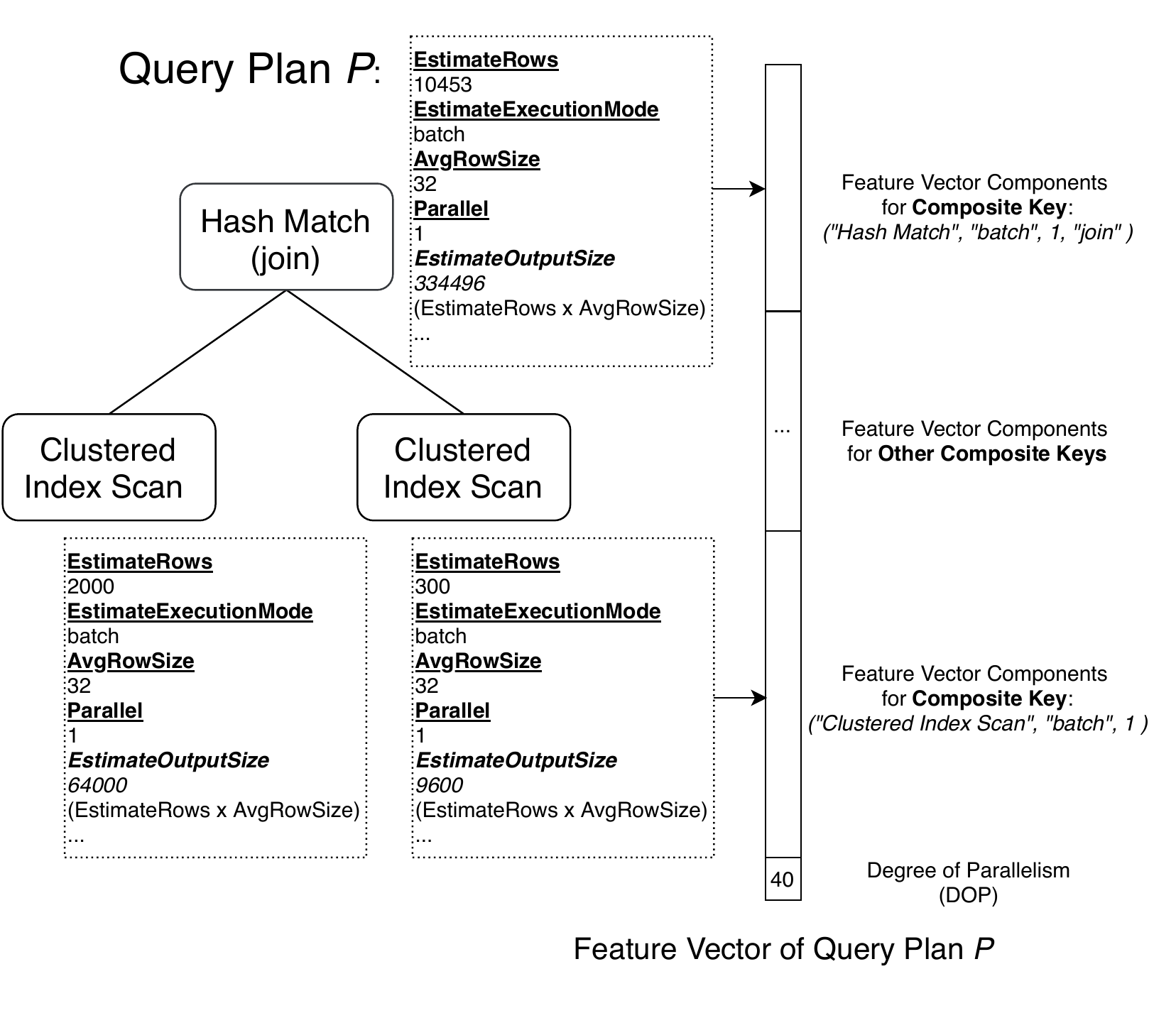}
\caption{Example of featurizing a query plan. The information of physical operators with different composite keys are encoded into different parts of the feature vector.}
\label{fig:plan_example}
\end{figure}

We next detail our featurization process. A simplified example of the process is depicted in Figure~\ref{fig:plan_example}.
Recall that $P$ is a tree, where each node in the tree consists of a physical operator along with its processing mode and information about its estimated runtime performance. Each physical operator can execute in different modes: $(i)$ \textsf{parallel} or \textsf{serial}, and $(ii)$ \textsf{batch} or \textsf{row}.
Since different modes handle parallelism differently, it is critical that we encode them in the feature vector. For example, it is important to distinguish between operators that use a parallel implementation versus the ones with a non-parallel implementation. 
To achieve this, we construct a composite key for each operator instance in the form of 
$$ (\mathsf{operator}, \mathsf{batch/row}, \mathsf{parallel/serial},  \langle \mathsf{optional} \rangle ) $$
The vector $\langle \mathsf{optional} \rangle$ adds additional attributes that are only applicable for certain operators. For example, \textsf{IsAdaptive} and \textsf{EstimatedJoinType} apply only for the operator \textsf{AdaptiveJoin}. In addition, in the case where the same physical operator can be used to implement different logical operators (e.g., \textsf{HashMatch} can be used for join, aggregation or union), $\langle \mathsf{optional} \rangle$ encodes the logical type as well.
 
For each composite key, the feature vector allocates space to encode different types of execution measures -- each measure is summed across all nodes in the plan with the same composite key. 
\begin{packed_item}
\item \textbf{Cardinality} (\fcard): cardinality estimates computed by the optimizer expressed in {\em bytes} (e.g., estimated output size in bytes).
\item \textbf{Cost} (\fcost): cost-based estimates calculated by the optimizer's cost model (e.g., estimated CPU cost, estimated I/O cost). 
\item \textbf{Count} (\fcount): takes value 1 if the operator has the composite key, otherwise 0.
\item \textbf{Weight} (\fweight): each node is assigned a weight which is computed recursively from the leaf nodes. The weight of a leaf node is the estimated output size in bytes, while for a non-leaf node it is the sum of product of weights and height of all its children. The weight feature encodes structural information (see~\cite{ding2019ai}).
\end{packed_item}

\subsection{ML Models}

We briefly describe the ML models we use in our approach.

\introparagraph{Linear Regression (LR)}
We consider Linear Regression as one of the baseline ML models due to its simplicity and efficiency. Specifically, we use its regularized version, {\em elastic net}, which linearly combines L1 and L2 regularization.

\introparagraph{Random Forest (RF)}
Random Forest is an ensemble learning method exploiting \textit{bagging} as its ensemble algorithm. Since multiple decision trees are built/traversed independently during the model construction/inference phase, this allows for efficient training/testing in a multicore setting. The fact that multiple trees built on different randomly sampled data also makes it more robust compared to a single decision tree, enabling it to consider complex interactions among different features while mitigating overfitting.

\introparagraph{XGBoost}
XGBoost \cite{chen2016xgboost} is an efficient implementation of gradient boosting with a set of optimizations that exploit any available parallelism, achieve faster model convergence by using the second-order gradients approach Newton Boosting and mitigate the overfitting issue of gradient boosting by imposing better regularization techniques.

\introparagraph{Multi-Layer Perceptron (MLP)}
The multi-layer perceptron is a class of feedforward artificial neural networks. In this work, we use a fully connected neural network with $8$ hidden layers in which each layer has $512$ neurons. RELU ($max(0, x)$) is used as the activation function. We use the Adam optimizer for training~\cite{kingma2014adam}.

\begin{figure}[ht]
	\subfloat[Training Convergence]{\includegraphics[width=0.25\textwidth]{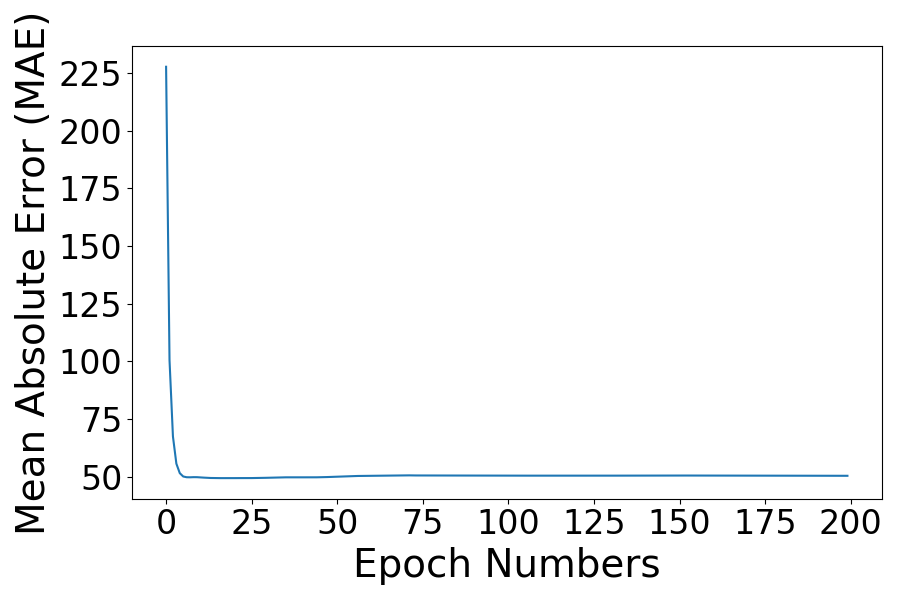}\label{fig:qpp_history}}
	\subfloat[Speedup Trend Prediction]{\includegraphics[width=0.25\textwidth]{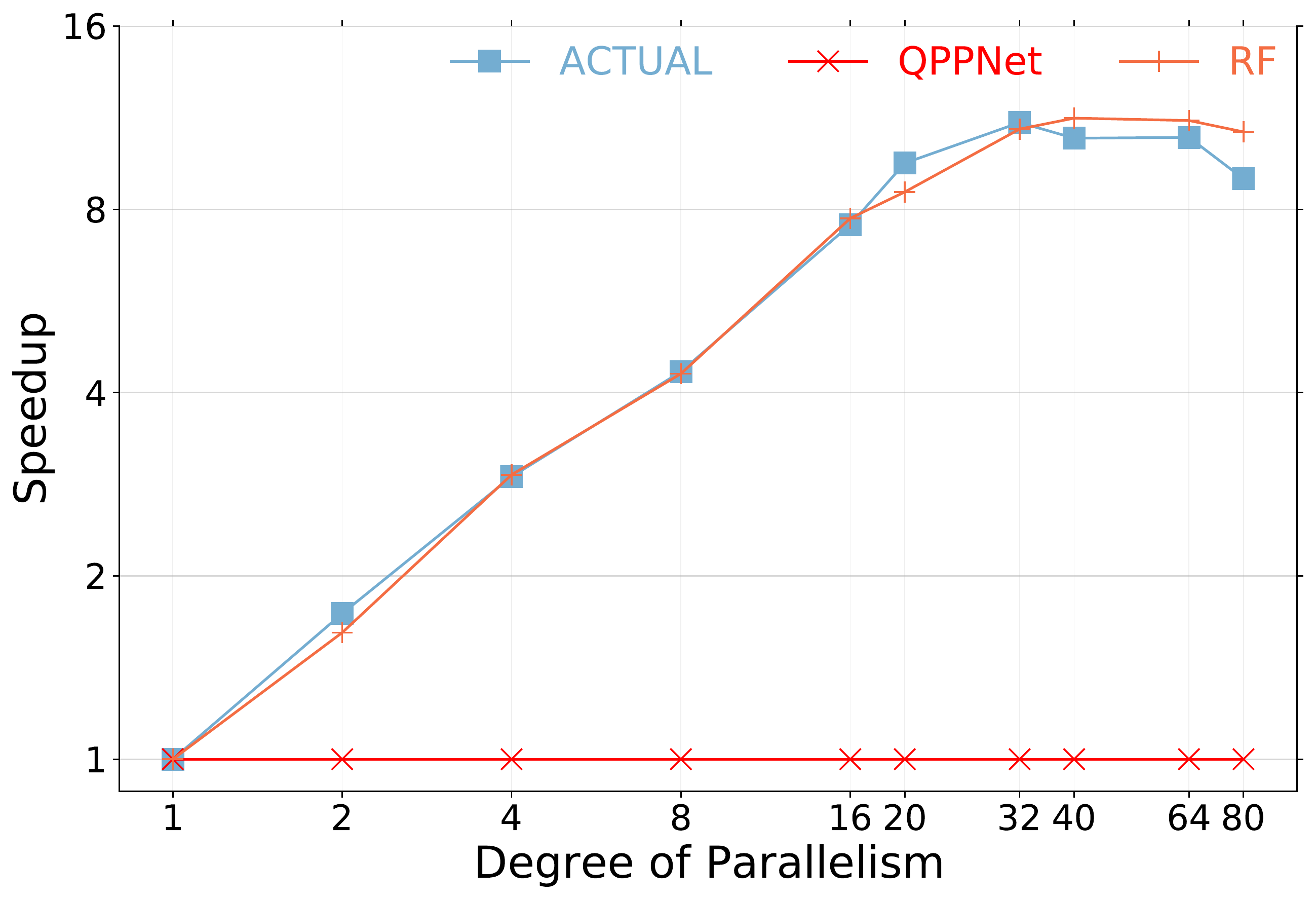}\label{fig:qpp_speedup}}
	\caption{\textbf{Example Result of QPPNet in QPP}: \ref{fig:qpp_history} shows QPPNet does converge after training for less than 25 epochs on 80\% of the TPC-DS1000 plan-dop pairs (\gl1). But \ref{fig:qpp_speedup} suggests that QPPNet is {\em ignorant} of different DOP values - it always gives the same latency prediction for the same query regardless of DOP changes, while RF is able to capture the actual speedup curve accurately.}
	\label{fig:qpp_net_comparison}
\end{figure}

\introparagraph{Exploration of Other Models}
Motivated by the promising results in~\cite{marcus2019plan} with explicit exploitation of the query plan structure and operator-level query latency, we investigated a rich set of DNN-based models that are designed to capture the spatial (e.g., nearby pixels in images) or structural (e.g., syntactic tree structure in natural languages) information of the data. Specifically, we explored models that have been applied successfully to a set of natural language processing (NLP) related tasks including convolutional neural network (CNN)~\cite{kim2014convolutional}, long short-term memory (LSTM) based recurrent neural network (RNN) \cite{sundermeyer2012lstm} and Tree-LSTM~\cite{tai2015improved}, regarding each operator instance as a {\em special word}. We observed that none of these models outperform MLP, the simplest DNN-based model, considering both runtime efficiency and prediction performance. We also implemented and explored QPPNet~\cite{marcus2019plan}, including the {\em operator-level elapsed time}. QPPNet failed to learn the relation between query latency and DOP (see example given in Figure~\ref{fig:qpp_net_comparison}). The failure could be attributed to a series of possible reasons, such as the difference of the studied system environment (e.g., PostgreSQL vs SQL Server ), the assumptions QPPNet relying on as discussed in the introduction, the not well-defined per-operator latency in our study environment, and the different sizes of query plans (more details in Figure~\ref{fig:operator-distrib}). Our models exploit the schema and database instance agnostic featurization, while QPPNet is tested on schema-dependent features.  Deeper investigation of the failure of applying QPPNet in DOP tuning is future work. Due to the above observations, we choose to use MLP as the DNN-based model representative and only present the comparison results of its with that of other models.

\begin{figure*}[ht]
\centering
\includegraphics[scale=0.5]{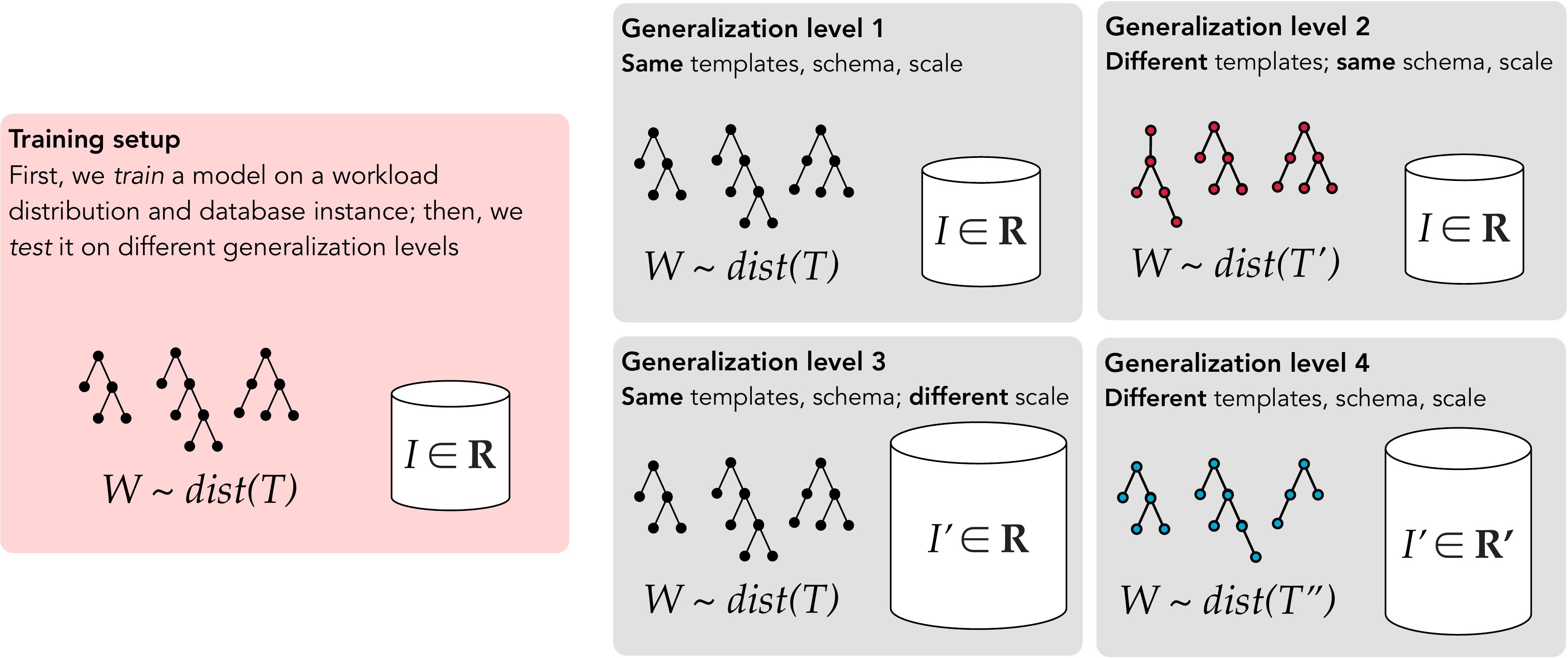}
    \caption{Overview of training setup and generalization levels. We use $\mathit{dist}(T)$ to denote a distribution of queries over a set of templates $T$.}
\label{fig:overview}
\end{figure*}

\subsection{Generalization} 
\label{subsec:generalization_levels}

Different application scenarios exhibit different \textit{degree of commonality}, i.e., how much similarity is observed across a query workload. The commonality here refers to both the queries themselves (e.g., SQL statements and query plans), as well as the input data. Given a specific application, having a good understanding of the degree of commonality is crucial when attempting to apply ML-based techniques for DOP tuning, since it helps practitioners to understand the capabilities and limitations of the models that are learned from the underlying data and queries. More specifically, {\em given training data (pairs of query plans plus runtime)}, we should first try to answer the following questions: 
{\em
\begin{packed_enum}
\item How similar are future queries to queries in the training data?
\item How much can the input relations change (in terms of both schema and scale) compared to the input relations in the training data?
\end{packed_enum} 
}

The answers to the above questions are different when considering different applications. For example, a supermarket chain might run the same set of queries daily to analyze its sales, and the size and the distribution of the input data to the queries might be relatively stable across most of the days, with the exception of a few  promotion days such as Black Friday and Christmas. On the other hand, analysts in different teams might run different queries on the same internal data for different data analysis tasks. In addition, the scale of the input data may increase after a period of time as new data arrives daily. 

To capture this differentiation, we categorize the \textit{training data} and \textit{test data} into four \textit{generalization levels}, considering the query templates, the schema of input tables and the scale of data. 
Suppose that we train our model on instance $I$ with schema $\mathbf{R}$, where the workload $W$ is drawn from a distribution of templates $dist(T)$ (see Figure~\ref{fig:overview}).


\introparagraph{\gl 1: \\same templates \& same schema \& same scale}\\
The test workload comes from the same distribution of templates $dist(T)$ and runs on the same database instance $I$. As an example, the training and test data are gathered from the executions of two sets of TPC-DS queries generated from the \textit{same set of templates} but with \textit{different random seeds} running against the \textit{same instance}.

\introparagraph{\gl 2: \\different templates \& same schema \& same scale}\\
The test workload comes from a different distribution of templates $dist(T')$, but runs on the same instance $I$. As an example, the training and test data are gathered from running two sets of TPC-DS queries generated from a \textit{different subset of templates} running against the \textit{same instance}.

\introparagraph{\gl 3: \\same template \& same schema \& different scale}\\
The test workload comes from the same distribution of templates $dist(T)$, but runs on an instance $I'$ of different scale using the same schema $\mathbf{R}$. As an example, the training and test data are gathered from running two sets of TPC-DS queries generated from \textit{the same set of query templates} running against the $300$ and $1000$ scale instances respectively.

\introparagraph{\gl 4: \\different template \& different schema}\\
The test workload comes from a different distribution of templates $dist(T'')$ and runs on a different instance $I'$ defined on a different schema $\mathbf{R'}$. As an example, the training and test data are gathered from running two sets of queries generated from TPC-DS templates and TPC-H templates running against the  corresponding database instances.

\smallskip
We assume a fixed hardware configuration and do not study generalization to different hardware configurations in this work. Cloud platforms generally have a restricted set of known hardware configurations on which services are deployed, and separate models could be trained and deployed for each configuration.

\section{Experimental Evaluation}
\label{sec:exp}

In this section, we present and discuss the results of our experimental evaluation. We design our experiments with the goal of answering the following questions:

\begin{itemize}[noitemsep,topsep=0pt]
\item What are the effects of different featurization alternatives in model performance?
\item How do different ML models perform for different generalization levels?
\item What are the root causes behind wrong predictions in the ML models in our problem formulation?
\item What trade-offs should be considered when choosing between different ML models? 
\end{itemize}

At a high level, our experiments show that the featurization using the \fcount, \fcard\ and \fweight\ features (excluding \fcost) leads to the overall best model prediction performance. RF shows the best overall performance considering both task-agnostic and task-specific metrics. 
However, when there is significant {distribution mismatch} between the test queries and training queries, no model is able to perform DOP selection that gives performance close to optimal. One concrete explanation for such distribution mismatch is the difference of memory requirements. 

Comparatively speaking, hyper-parameter tuning of  XGBoost and MLP is time-consuming while additional regularization is hardly to be imposed for better generalization (e.g., to achieve better performance on test data) when the training data is limited; RF is easier to be configured and it is more robust while it has relatively higher model inference overhead. 

\subsection{Experimental Setup} 
\label{subsec:experimental-setup}

\introparagraph{System Configuration}
Our queries were executed on a dual-socket Intel Xeon Broadwell server with a total of $40$ physical cores and $512$ GB main memory. We have hyper-threading enabled, resulting in a total of $80$ logical cores which, consequently, is the maximum DOP possible on this server.
 We run SQL Server 2019 CTP 2.0 Developer edition. By default, SQL Server chooses a DOP value of 64 for queries running on this server. SQL Server sets aside part of the available memory for the buffer pool and shared objects; the remainder can be granted as working memory to queries up to a certain limit (by default, 25\% of the working memory, corresponding to $\sim$90GB on our server). Using query hints, we explicitly request this maximum memory grant for queries that spill. The data for the databases we study are striped across several NVMe SSDs and their logs files are on another NVMe SSD. 

\introparagraph{Workload and Datasets}
We use both the TPC-H and TPC-DS workloads for our experiments, with scale factors 100, 300, and 1000 for each. The following table gives the detailed information for the dataset construction.

\begin{table}[H]
\caption{Workload and Dataset Statistics}
\setlength\tabcolsep{3.0pt}
\begin{tabular}{  l | c | c | c }  
\toprule
\textbf{} &\textbf{\#templates} & \textbf{\#queries} & \textbf{\#data points} \\
\midrule
\textbf{TPC-H 100} & 22 & 1346 & 13460 \\
\textbf{TPC-H 300} & 22 & 1346 & 13460 \\
\textbf{TPC-H 1000} & 22 & 1346 & 13460 \\
\textbf{TPC-DS 100} & 103 & 841 & 8410 \\
\textbf{TPC-DS 300} & 106 & 853 & 8530 \\
\textbf{TPC-DS 1000} & 103 & 841 & 8410 \\
%
%
%
\bottomrule
\end{tabular}
\label{tab:meta_data}
\end{table}

For each TPC-H query template, we generated queries using $100$ different random seeds with duplicate queries being removed. TPC-DS queries are generated in a similar manner, but with $10$ different random seeds used for each query template.
We execute each query in the workloads for $10$ different DOP values: $\{ 1, 2, 4, 8, 16, 20, 32, 40, 64, 80\}$.
We focus on warm-cache query execution, since caching is almost always used in practice whenever it is feasible. Thus, to obtain each data point, we perform one cold-cache query execution followed by ten warm-cache runs. The average time of the warm-cache runs is used as the ground truth of the latency for the executed query. For the purpose of our study, we execute one query at a time. We also only consider queries executed using query plans with at least one physical operator that uses intra-parallelism.
We use the clustered columnstore index~\cite{cci} organization for each database. 

\introparagraph{Training and Testing Splits}
We perform 5-fold cross-validation to evaluate the model performance for each generalization level. 
For \gl 1, each fold uses $80\%$ of TPC-DS 1000 queries for training and the remaining $20\%$ for testing. 
For \gl 2, the data is split based on the query templates: each fold uses $80\%$ of the query templates from TPC-DS1000 for training, and the remaining $20\%$ for testing. The folds for \gl 3 and \gl 4 are as follows:
 
\begin{table}[H]
\caption{TPC Queries in Training/Testing Folds for \gl 3  \text{ and} \gl 4.}
\begin{center}
\setlength\tabcolsep{3.0pt}
\begin{tabular}{  l | c | c }  
\toprule
\textbf{} & \textbf{\gl 3} & \textbf{\gl 4} \\
\midrule
\textbf{Fold 1} & DS1000/DS300 & DS1000/H1000 \\
\textbf{Fold 2} & DS1000/DS100 & DS1000/H300 \\
\textbf{Fold 3} & DS300/DS100 & DS1000/H100 \\
\textbf{Fold 4} & DS100/DS300 & DS300/H1000 \\
\textbf{Fold 5} & DS100/DS1000 & DS300/H300 \\
\bottomrule
\end{tabular}
\label{tab:gen_3_gen_4_train_test}
 \label{tab:gen_3_gen_4_train_test}
\end{center}
\end{table}
\introparagraph{Hyper-Parameter Tuning}
We tune the selected models in a standard manner, similar to \cite{ding2019ai}, based on the mean absolute error in cross-validation on the training data. For tree-ensemble models (XGBoost and RF), we limit the maximal number of trees to be $1000$. We note that spending additional effort tuning the parameters of different models might lead to a narrower gap between the evaluation metrics of these models. However, the main goal of this study is to
to explore the usability and trade-offs present in the compared techniques in a standard training configuration. Hence, for each model, we stop tuning the hyper-parameters (e.g., searching larger hyper-parameter space) when reasonably satisfying results are observed. 

\introparagraph{Comparison Metrics}
To compare the performance of different ML models, we use both {\em task-agnostic} metrics (e.g., mean absolute error), as well as {\em task-specific} metrics that are tied to a specific application (e.g., workload throughput at optimal DOP selection). We provide the detailed definitions of these metrics in Table~\ref{tab:comparison_metrics}.


\begin{table*}[t]
    \caption{Comparison Metrics}
        \label{tab:comparison_metrics}
\centering
        \begin{tabular}{ l | l | l  }
        	   \toprule
	   \textbf{Category} & \textbf{Metric} &\textbf{Definition} \\
            \midrule
            
            \multirow{4}{*}{Task-Agnostic} 
            			&  Mean Absolute Error &  $\mathsf{MAE}(W) = \frac{1}{|{W}| |\mathbf{D}|}\sum_{i=1}^n \sum_{d \in \mathbf{D}}{|\hat{t}_{(P_i, d)} - t_{(P_i, d)}|}$\\
             			& Relative Prediction Error &  $\mathsf{RPE}(P) = \frac{1}{|\mathbf{D}|}\sum_{d \in \mathbf{D}}\frac{|\hat{t}_{(P, d)} - t_{(P, d)}|}{t_{(P, d)}}$ \\
                                 & Speedup Prediction Error &  $\mathsf{SPE}(P) = \frac{1}{|\mathbf{D}|}\sum_{d \in \mathbf{D}}|\frac{\hat{t}_{(P, d)}}{\hat{t}_{(P, 1)}} - \frac{t_{(P, d)}}{t_{(P, 1)}}|$ \\
                   
            \midrule    
            \multirow{2}{*}{Task-Specific} 
            		       & Throughput with per-Query DOP & $\mathsf{TQ}(W) =  \frac{|W| }{ \sum_{i=1}^n \min_{d \in \mathbf{D}} {\hat{t}_{(P_i, d)}}}$\\
		       	      & Throughput with per-Workload DOP &  $\mathsf{TW}(W) =  \frac{|W| }{ \min_{d \in \mathbf{D}} \{ \sum_{i=1}^n {\hat{t}_{(P_i, d)}} \}}$\\
	   \bottomrule
        \end{tabular}
\end{table*}

 \vspace{-4pt}
\subsection{Featurization Analysis}
\label{sec:featurization_analysis}

To compare the featurization alternatives, we run RF with all four features $F = \{\fcard, \fcost, \fcount, \fweight\}$, and also excluding one feature at a time ($F\backslash \{\fcard\}$, $F\backslash \{\fcost\}$, $F\backslash \{\fcount\}$, $F\backslash \{\fweight\}$).
 Tables \ref{tab:featurization_rpe_g1}-\ref{tab:featurization_spe_g4} show the distributions of relative prediction error (RPE) and speedup prediction error (SPE). 
We observe that for \gl 1 there is little difference across different featurization alternatives (Tables \ref{tab:featurization_rpe_g1},\ref{tab:featurization_spe_g1}). For the other generalization levels, $F\backslash \{\fcost\}$ leads to the best performance: for the  {low relative error range} (e.g., RPE $<$ 0.80), the  percentage given by using $F\backslash \{\fcost\}$ is larger than other featurization alternatives.
We also observe that including all features $F$ is rarely optimal, while having \fcount\ and \fweight\  is beneficial most of the time. 
As a result of our analysis, we  present our experimental results using the features $\{ \fcount, \fcard, \fweight\}$ for the remaining experiments.


\begin{table*}
\begin{center}
\caption{5-Fold Mean Absolute Error on Training Data}
\setlength\tabcolsep{2.0pt}
\begin{tabular}{ c  c  c  c  c } 
\toprule
\textbf{Model} & \textbf{\gl 1} & \textbf{\gl 2} & \textbf{\gl 3} & \textbf{\gl 4}  \\ 
\midrule

\textbf{LR} &{67.6/63.8/67.9/67.5/64.6} & {66.6/60.9/57.1/61.8/62.7} & {64.9/64.6/28.4/9.6/9.6} & {64.5/64.2/64.2/28.5/28.4} \\

\textbf{MLP} & {3.7/4.2/3.2/3.7/10.8} & {2.6/3.2/3.1/3.3/4.9} & {3.5/5.5/5.1/0.8/0.5} & {3.7/3.3/5.0/1.3/2.0}  \\

\textbf{RF} & {2.1/2.0/1.9/2.1/2.0} & {1.7/1.3/1.2/1.4/1.3} & {1.8/1.8/0.6/0.3/0.3}  & {1.8/1.8/1.8/0.6/0.6}   \\

\textbf{XGBoost} & {0.6/0.6/0.3/0.6/0.4} & {0.5/0.2/0.4/0.4/0.4} & {0.5/0.5/0.3/0.1/0.1} & {0.5/0.5/0.5/0.3/0.3}  \\
\bottomrule
\end{tabular}
\label{tab:training_mae}
\end{center}
\end{table*}

\subsection{Training and Inference Time}

In this section, we evaluate the training and inference overheads for the ML models we tested. It is critical that both overheads are small for the model to be usable in practice. We report below (Table~\ref{tab:inference_time}) the training and inference overheads for \gl 1. 

\begin{table}[H]
\begin{center}
\caption{Model overheads for \gl 1.}
\setlength\tabcolsep{3.0pt}
\begin{tabular}{  l |  c |  c   } 
\toprule
\textbf{Model} & \textbf{Training (sec)} & \textbf{Inference (sec)} \\
\midrule

\textbf{LR} & {3} & {0.000349} \\

\textbf{MLP} & {81} & {0.009772} \\

\textbf{RF} & {17} & {0.419193} \\

\textbf{XGBoost} & {23} & {0.0013} \\
\bottomrule 
\end{tabular}
\label{tab:inference_time}
\end{center}
\end{table}

The training overhead includes the time spent on preprocessing (e.g., featurization, training/test data splitting, etc). MLP is trained on NVIDIA GeForce GTX 1080Ti, while other models are trained on the same machine in which the queries are executed with all $40$ physical cores given. The inference time of all models is measured by running each model on a single data point using a single CPU core. We observe that the inference time of all models except for RF is very small ($<$ 10ms ), while the training time varies, with MLP being the most expensive model. RF spent more time ($\sim400$ ms) on inference due to the traversal of large number of deep trees. However, our work focus on relatively long-running queries (e.g., OLAP), for which DOP tuning is mostly beneficial to, and thus we see the inference overhead of RF as insignificant. 


\vspace{-4pt}
\subsection{Model Comparison}
\vspace{-4pt}
\subsubsection{Task-Agnostic Metrics}

\introparagraph{Mean Absolute Error}
MAE represents the average absolute difference between the actual query latency and predicted query latency of all plan-dop pairs. A nice property of MAE is that it has the same unit as the regression target \textit{time}, which is also noted in \cite{marcus2019plan}. 

In Table \ref{tab:training_mae}, we observe that except for LR, all other models exhibit relatively small MAEs on the training data after learning across all generalization levels, suggesting the necessity of nonlinearity and large model capacity. Among these models, XGBoost and RF show the best performance considering only MAEs on the training data, which are close to zero, suggesting the number/depth of the trees in tree-ensemble models  
are large/deep enough.

Switching to test data, the comparison between results of LR and other models in \gl 1 are relatively consistent with what is being observed in training data. However, we see much larger gaps across different generalization levels in all models, and this suggests the different degree of difficulty of applying ML models for DOP tuning at different generalization levels. 
MAEs of different testing folds in each generalization level suggest that RF generalizes better than XGBoost and MLP. This result can be explained by the fact that RF is less sensitive to overfitting compared to XGBoost and MLP, as well as easier to tune.
While it is possible to apply better regularization techniques on XGBoost and MLP to further reduce the generalization error, we note that the current architecture/hyperparameters of XGBoost, MLP, and RF (which are tuned via cross-validation on the training data) show comparable performance on different validation splits.
The generalization error gap between different models might be narrowed down when 
more training data from queries of larger variety becomes available.


\begin{figure*}[!t]
\centering
\includegraphics[width=0.6\textwidth]{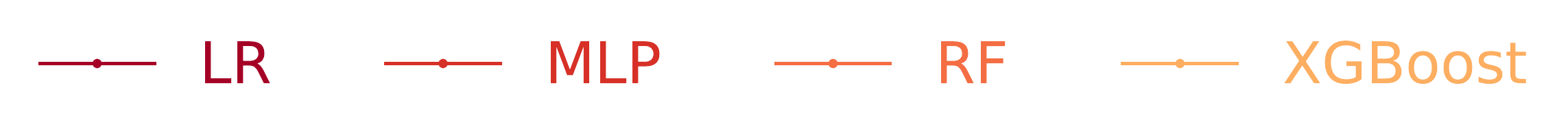}
\vskip-5ex
		\subfloat[\gl 1]{\includegraphics[width=0.24\textwidth]{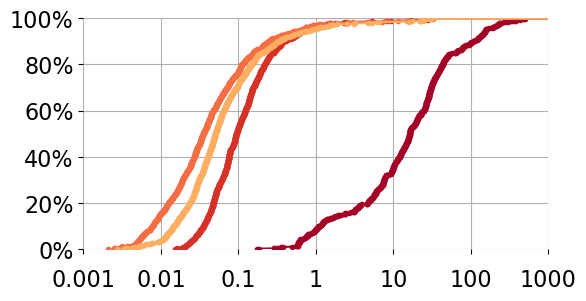}}
		\subfloat[\gl 2]{\includegraphics[width=0.24\textwidth]{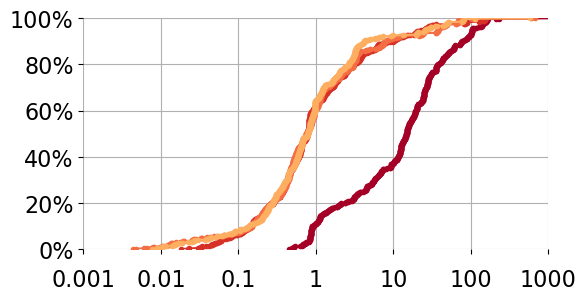}}
		\subfloat[\gl 3]{\includegraphics[width=0.24\textwidth]{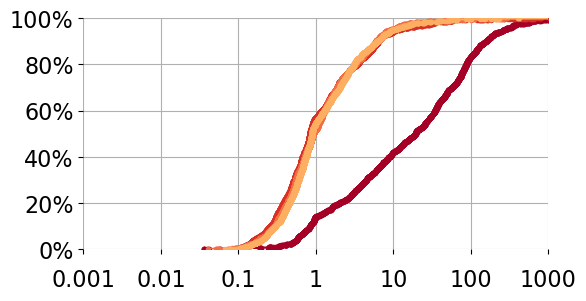}}
		\subfloat[\gl 4]{\includegraphics[width=0.24\textwidth]{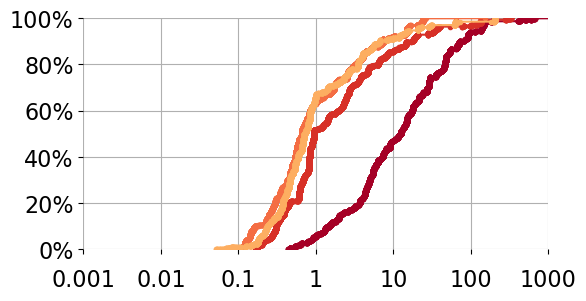}}
	\caption{Relative Prediction Error Distribution --- Percentage (Y-Axis)  vs. RPE (X-Axis)}
	\label{fig:relative_error_distribution}
		\subfloat[\gl 1]{\includegraphics[width=0.24\textwidth]{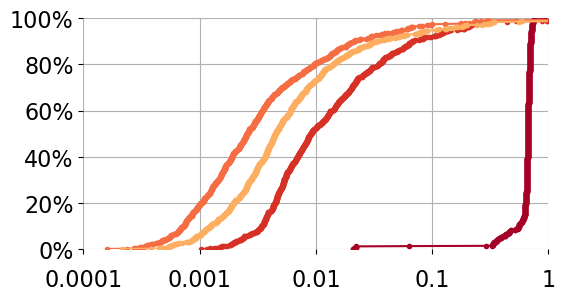}}
		\subfloat[\gl 2]{\includegraphics[width=0.24\textwidth]{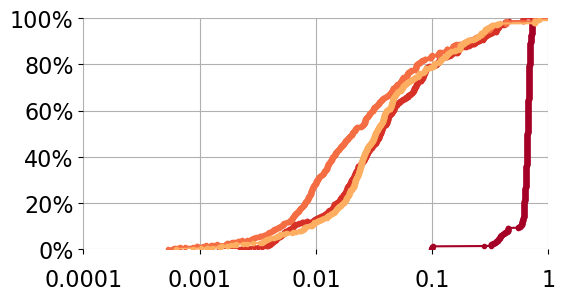}}
		\subfloat[\gl 3]{\includegraphics[width=0.24\textwidth]{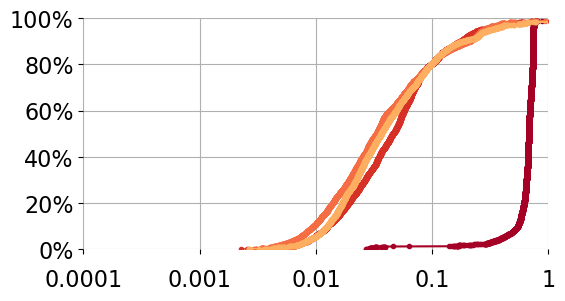}\label{fig:gen3_spe}}
		\subfloat[\gl 4]{\includegraphics[width=0.24\textwidth]{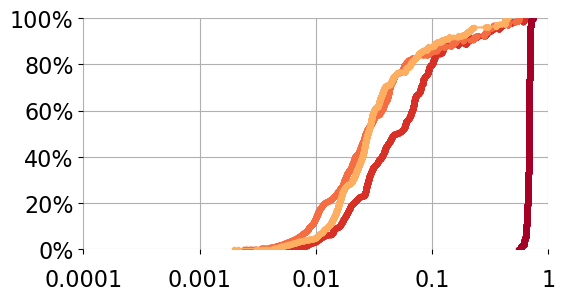}\label{fig:gen4_spe}}
	\caption{Speedup Prediction Error Distribution --- Percentage (Y-Axis) vs. SPE (X-Axis) }
	\label{fig:speed_up_error_distribution}
\end{figure*}

\introparagraph{Relative Prediction Error}
We show the distribution of the per-query \textsf{RPE}  of each model at different generalization levels in Figure \ref{fig:relative_error_distribution}. Not surprisingly, LR consistently exhibits high RPE ($> 0.5$) in most of the plan-dop pairs  across all generalization levels due to its inability of capturing the complex relationship between the plan-dop pair features and the query latency. Looking at the relative error distribution of other models, we observe that in most cases, RF exhibits lower RPE for a larger portion of test plan-dop pairs compared to other models.  

\introparagraph{Speedup Prediction Error}
While relative error might be a good metric for evaluating the accuracy of the query latency prediction, it does not directly infer the {\em speedup} of the query execution at different DOP values. Intuitively, when selecting the DOP for a single query execution, users should know the pattern or trend of the performance curve of the query at DOP values of interests. We look at a simple metric called {\em speedup prediction error} (Table \ref{tab:comparison_metrics}) that captures this property. Figure~\ref{fig:speed_up_error_distribution} presents the per-query based SPE distribution of each model considering different generalization levels. We observe that RF consistently shows the best performance for SPE. 

\begin{figure*}[!t]
\centering
\includegraphics[width=0.6\textwidth]{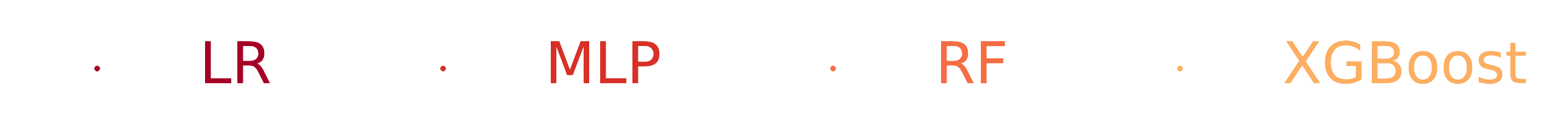}
\vskip-3ex
		\subfloat[\gl 1]{\includegraphics[width=0.25\textwidth]{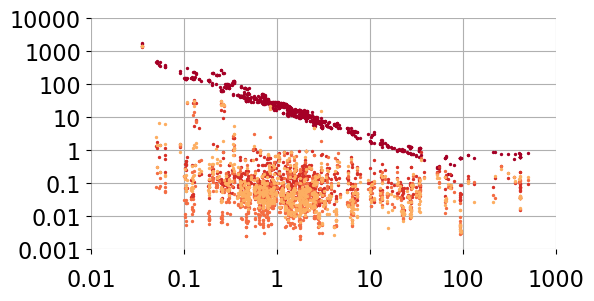}}
		\subfloat[\gl 2]{\includegraphics[width=0.25\textwidth]{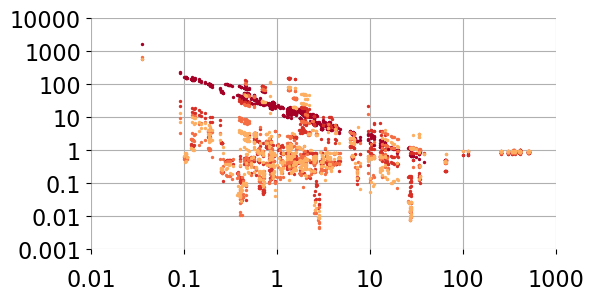}}
		\subfloat[\gl 3]{\includegraphics[width=0.25\textwidth]{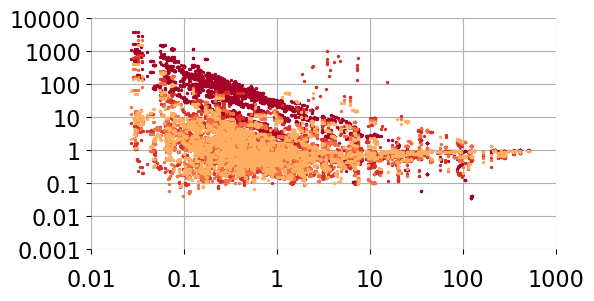}}
		\subfloat[\gl 4]{\includegraphics[width=0.25\textwidth]{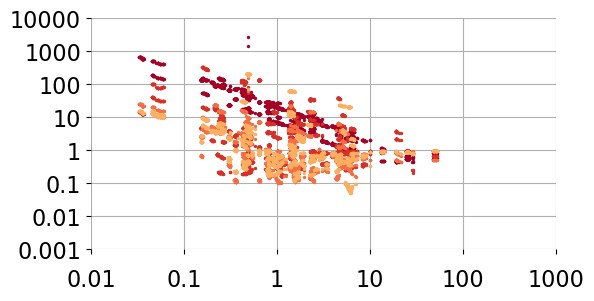}}
	\caption{RPE (Y-axis) vs. Query Latency at DOP 40 (X-axis)}
	\label{fig:relative_error_vs_time}
%
		\subfloat[\gl 1]{\includegraphics[width=0.25\textwidth]{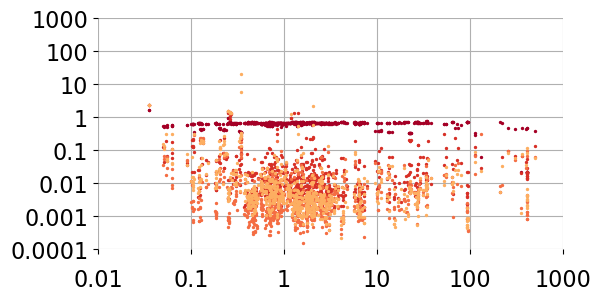}}
		\subfloat[\gl 2]{\includegraphics[width=0.25\textwidth]{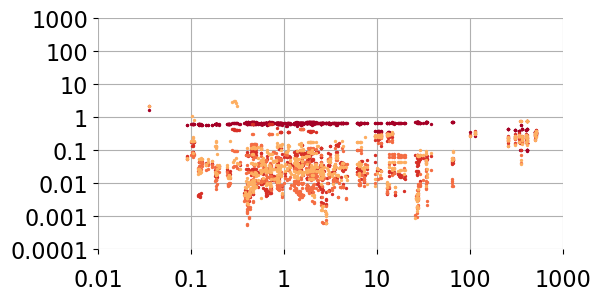}}
		\subfloat[\gl 3]{\includegraphics[width=0.25\textwidth]{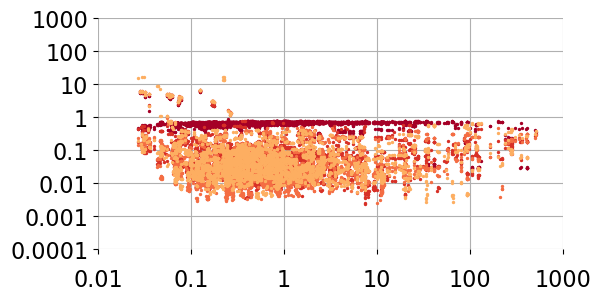}}
		\subfloat[\gl 4]{\includegraphics[width=0.25\textwidth]{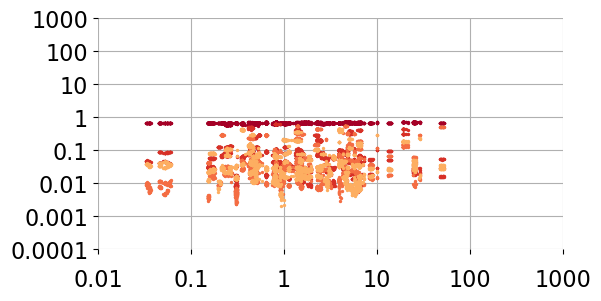}}
	\caption{SPE (Y-axis) vs. Query Latency at DOP 40 (X-axis)}
	\label{fig:speedup_error_vs_time}
\end{figure*}

The relationships between relative error/speedup error and the query latency (at DOP 40) are shown in Figure 
\ref{fig:relative_error_vs_time} and \ref{fig:speedup_error_vs_time}. We observe that large errors are less likely to occur in long-running queries ($>1s$) for both metrics. The observation is positive, since intra-parallelism is  more beneficial to long running queries.

\begin{table*}
\begin{center}
\setlength\tabcolsep{2.0pt}
\caption{5-Fold Mean Absolute Error on Test Data}
\begin{tabular}{  c  c  c  c  c } 
\toprule
\textbf{Model} & \textbf{\gl 1} & \textbf{\gl 2} & \textbf{\gl 3} & \textbf{\gl 4}  \\ 
\midrule
\textbf{LR} & {57.0/64.6/56.6/58.4/63.3} & {58.2/75.4/86.4/70.8/64.5} & {49.4/43.1/18.6/21.5/46.6} & {41.8/37.6/39.7/28.3/14.9} \\

\textbf{MLP} & {4.5/4.4/3.6/4.0/8.9} & {30.4/63.4/64.4/61.8/107.7} & {28.8/28.0/5.9/12.4/155.9}& {133.7/74.2/15.2/103.7/30.6}  \\

\textbf{RF} & {3.0/3.5/2.5/2.2/2.6} & {41.9/83.7/51.7/43.6/78.0} &{21.8/21.1/12.0/14.8/40.4} & {25.2/7.6/4.6/21.4/6.0}  \\

\textbf{XGBoost} & {4.0/2.9/2.9/2.4/5.9} & {31.1/66.4/64.1/62.1/106.6} & {26.9/27.7/12.1/15.2/41.3} & {65.7/49.0/47.0/22.8/7.8}  \\

\bottomrule
\end{tabular}
\label{tab:testing_loss}
\end{center}
\end{table*}


\begin{figure*}[!t]
\centering
\includegraphics[width=0.8\textwidth]{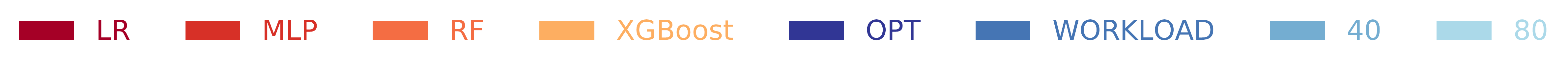}
\vskip-3ex
	\subfloat[\gl 1]{\includegraphics[width=0.50\textwidth]{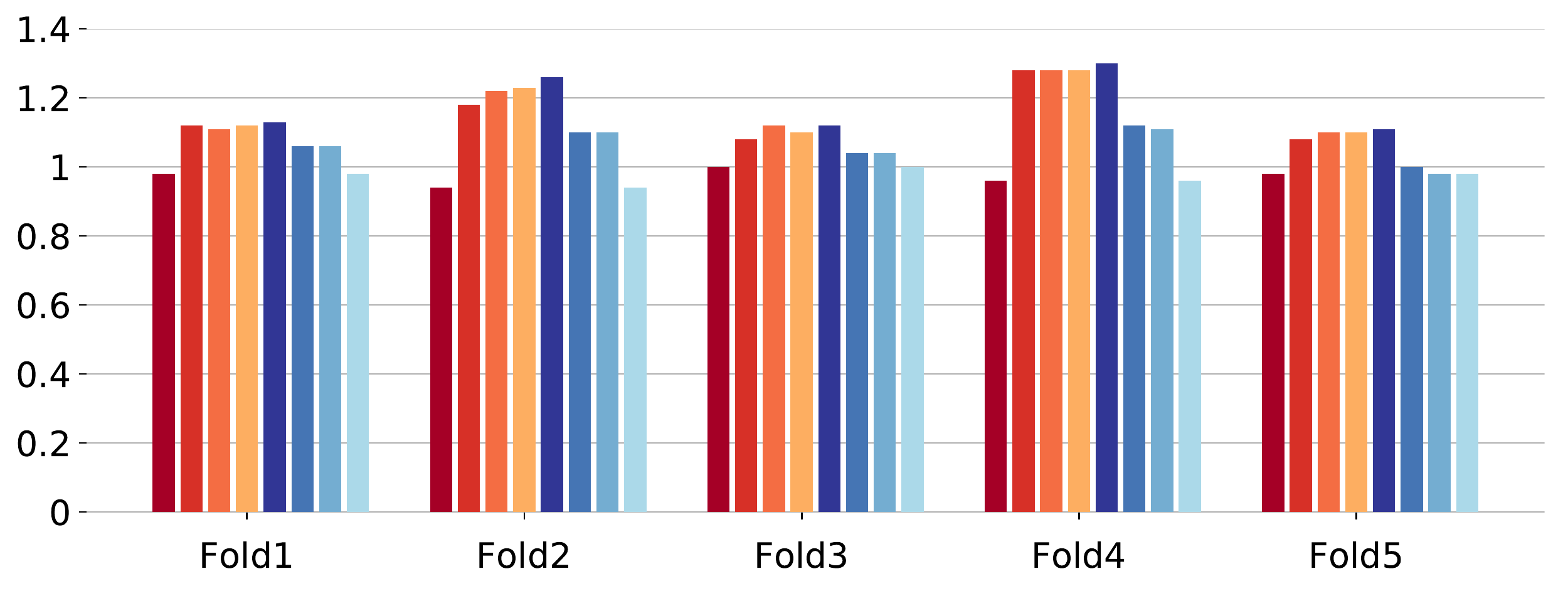}}
	\subfloat[\gl 2]{\includegraphics[width=0.50\textwidth]{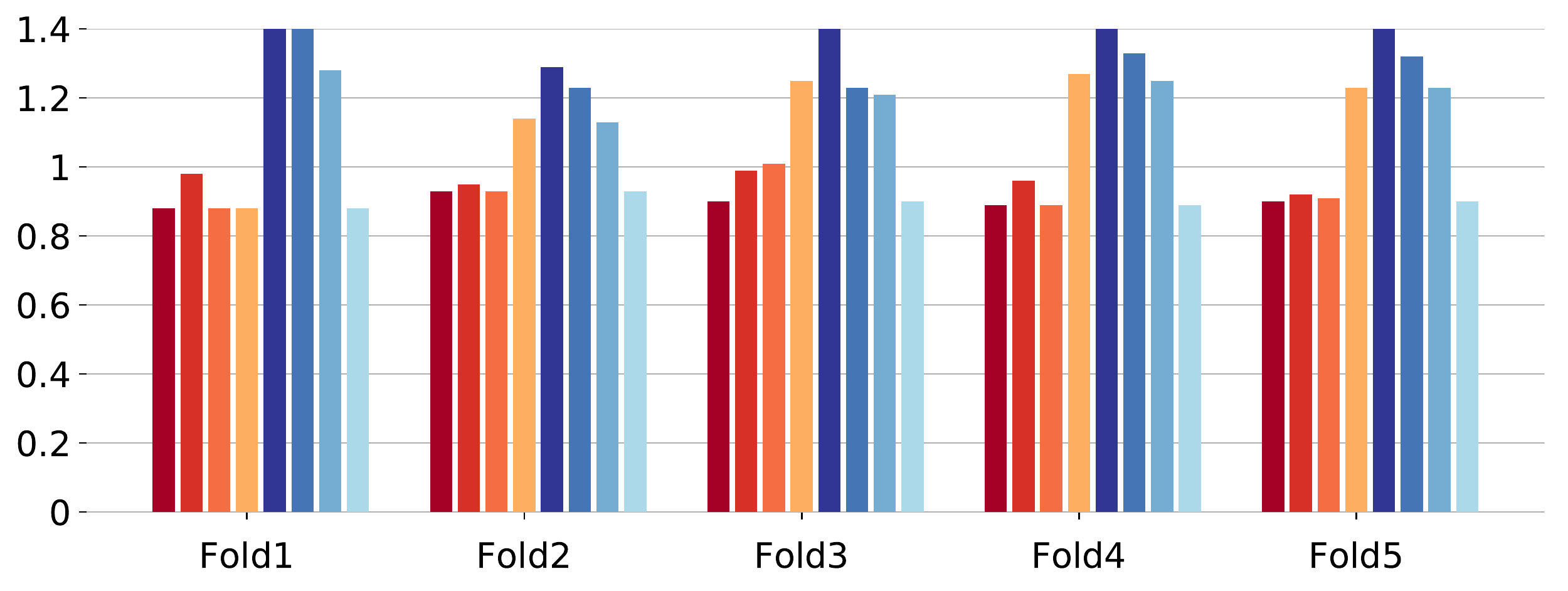}} \\\vspace{-10pt}
	\subfloat[\gl 3]{\includegraphics[width=0.50\textwidth]{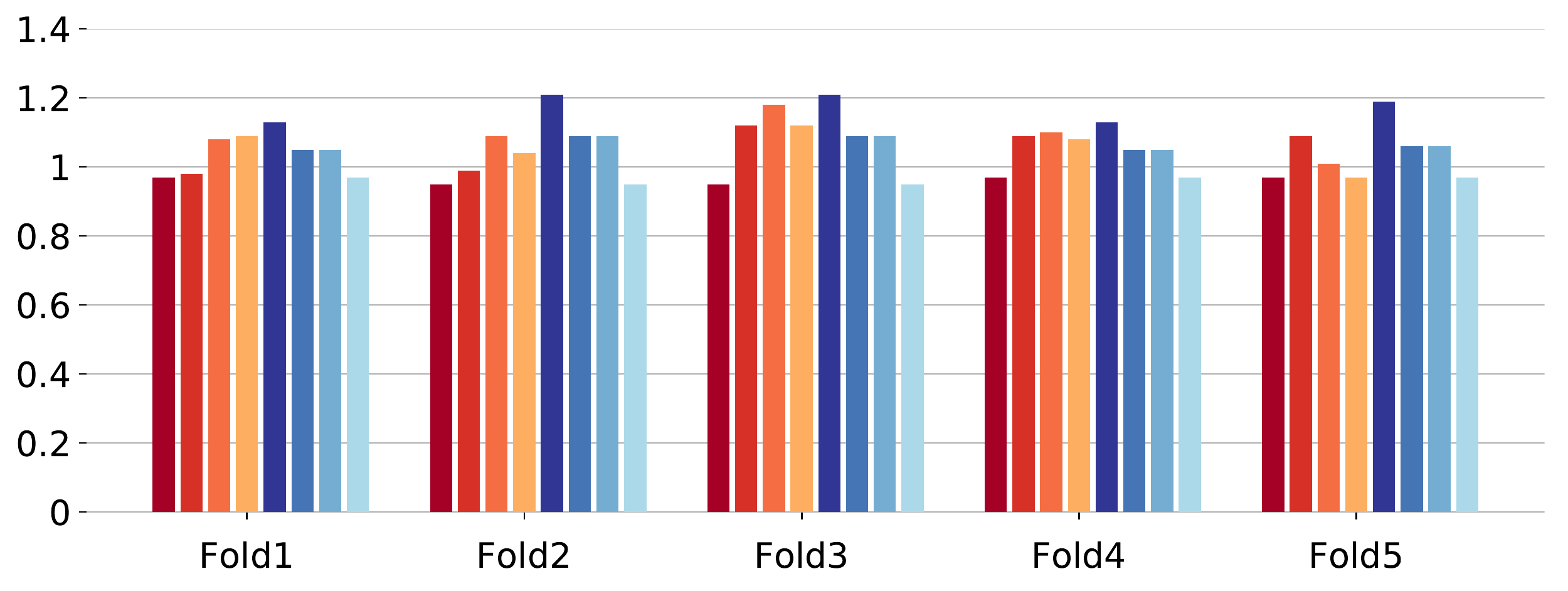}}
	\subfloat[\gl 4]{\includegraphics[width=0.50\textwidth]{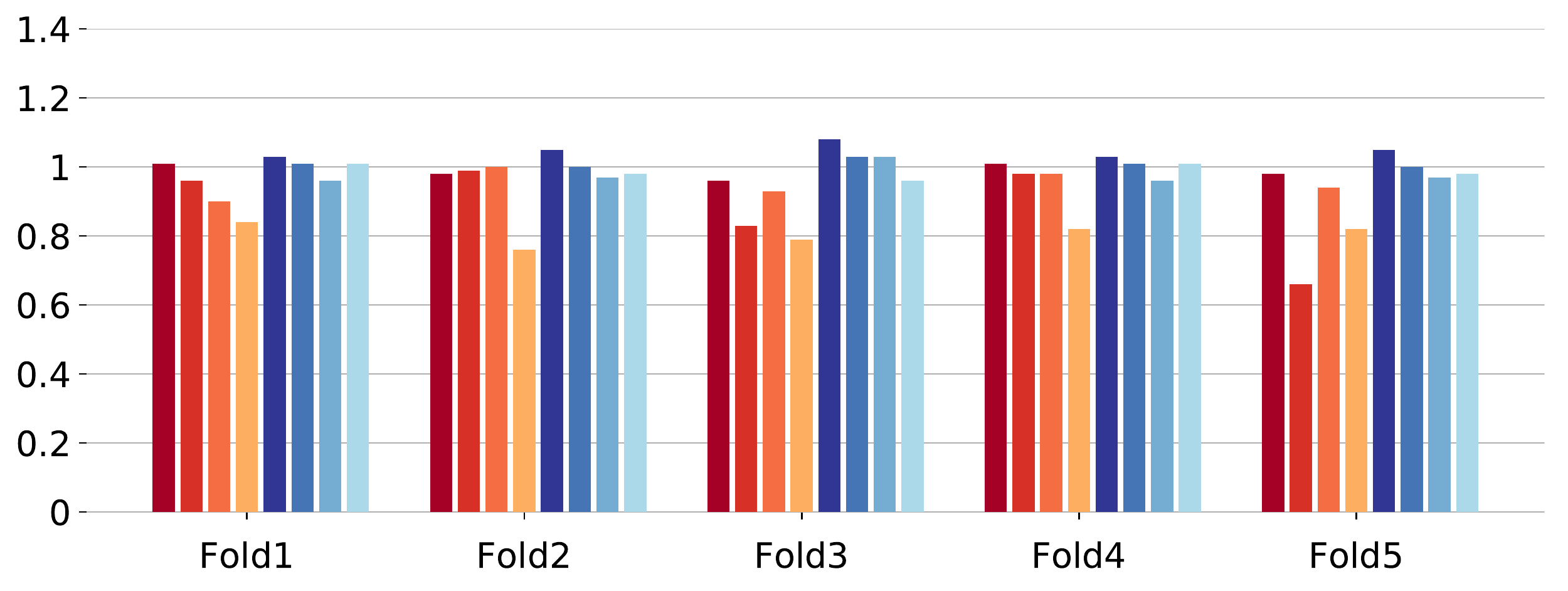}} 
	\caption{DOP Selection at Individual Query Level --- Query Throughput over DOP 64 (Y-Axis)}
	\label{fig:individual_dop_selection}
\end{figure*}
\begin{figure}
	\centering
	{\includegraphics[width=0.50\textwidth]{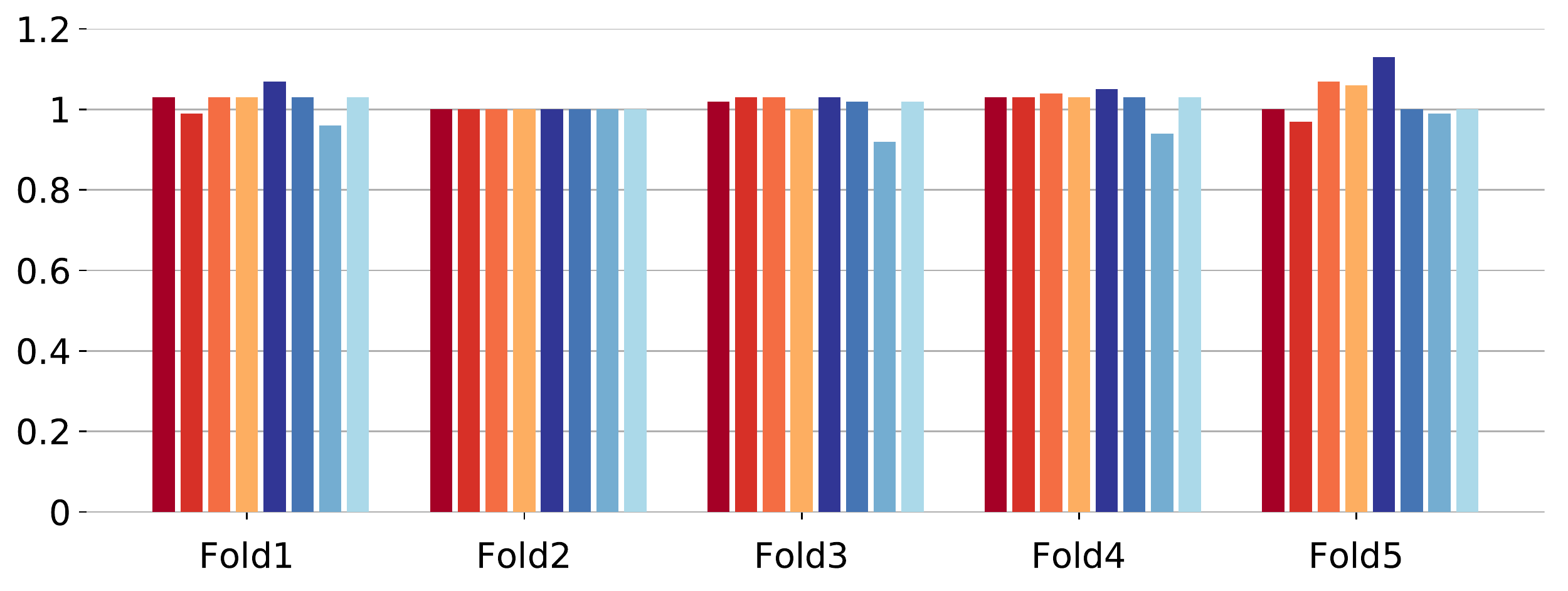}}
	\caption{\gl 2M}
	\label{fig:gen_2_mem}
\end{figure}

\begin{figure*}[!t]
\centering
\includegraphics[width=0.6\textwidth]{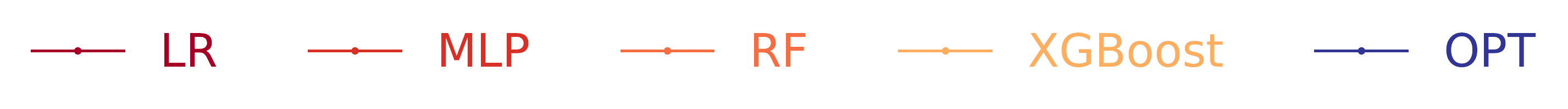}
\vskip-3ex
		\subfloat[\gl 1]{\includegraphics[width=0.20\textwidth]{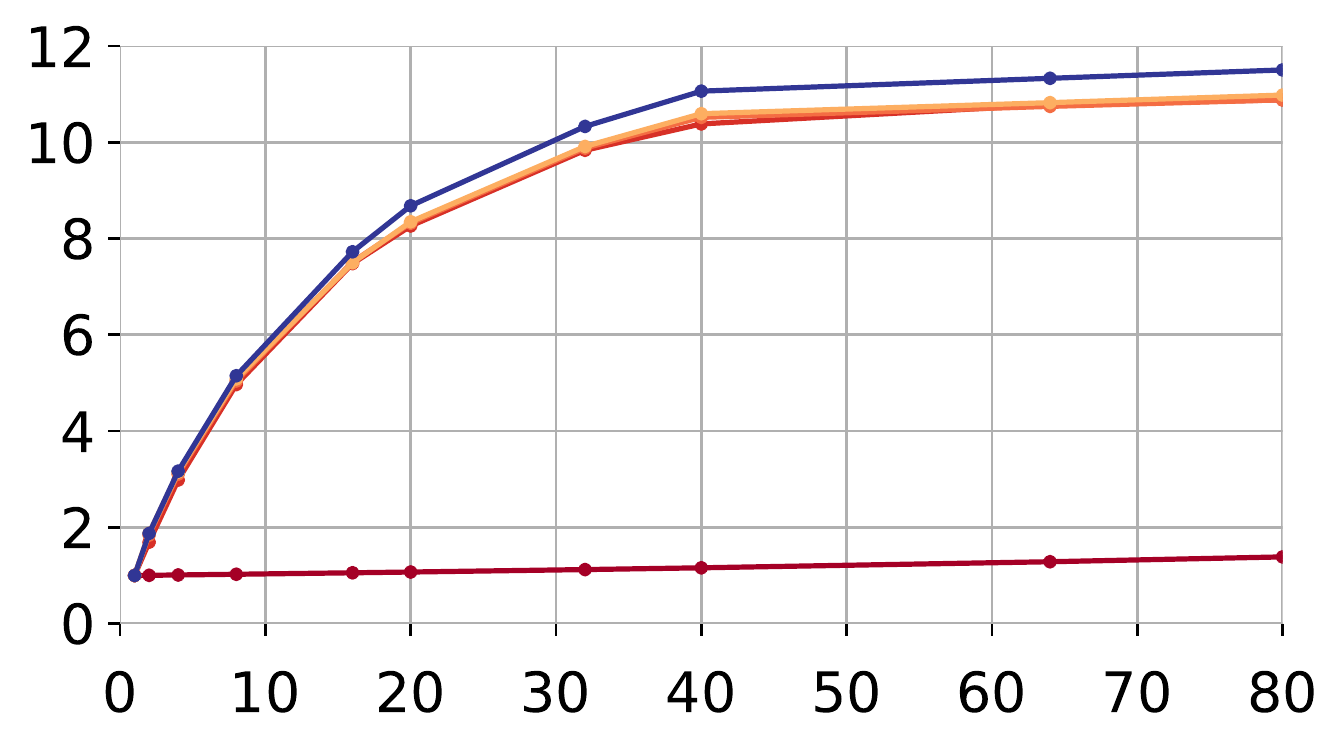}\label{fig:g1_dop_prediction}}
		\subfloat[\gl 2]{\includegraphics[width=0.20\textwidth]{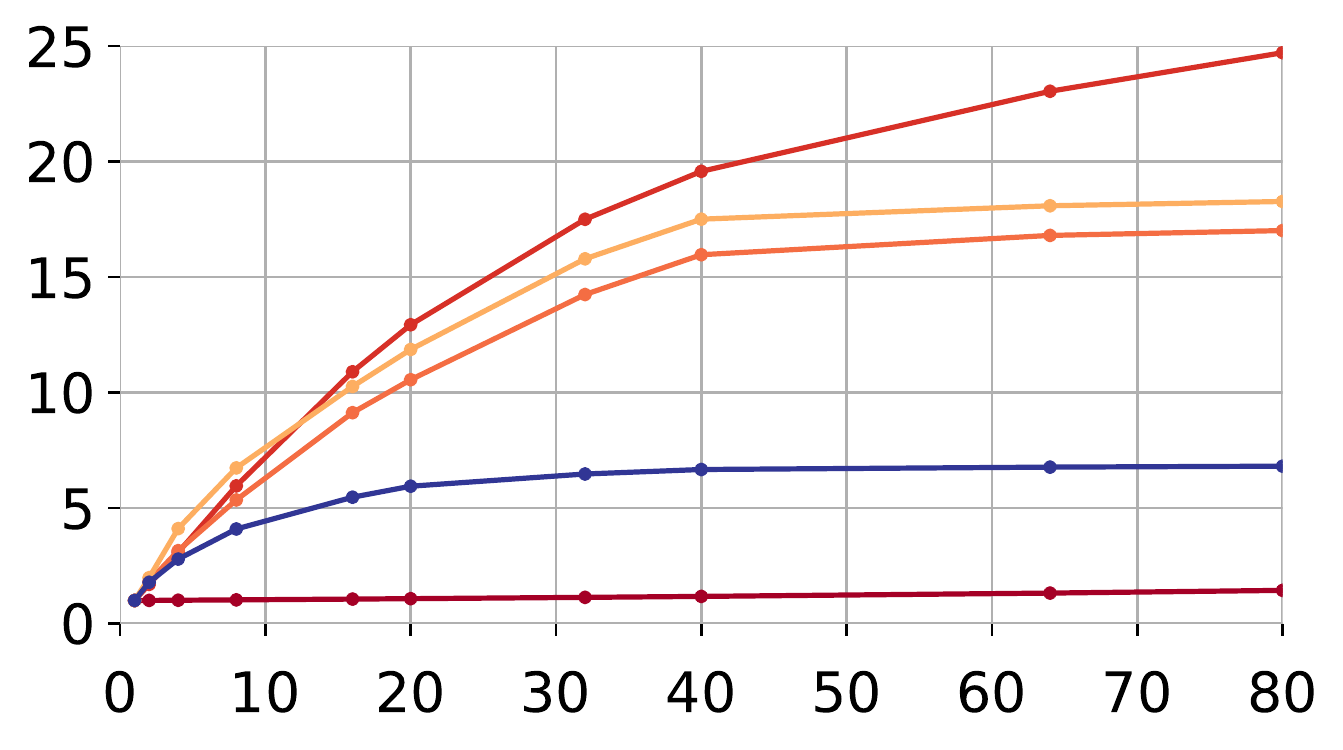}\label{fig:g2_dop_prediction}}
		\subfloat[\gl2M]{\includegraphics[width=0.20\textwidth]{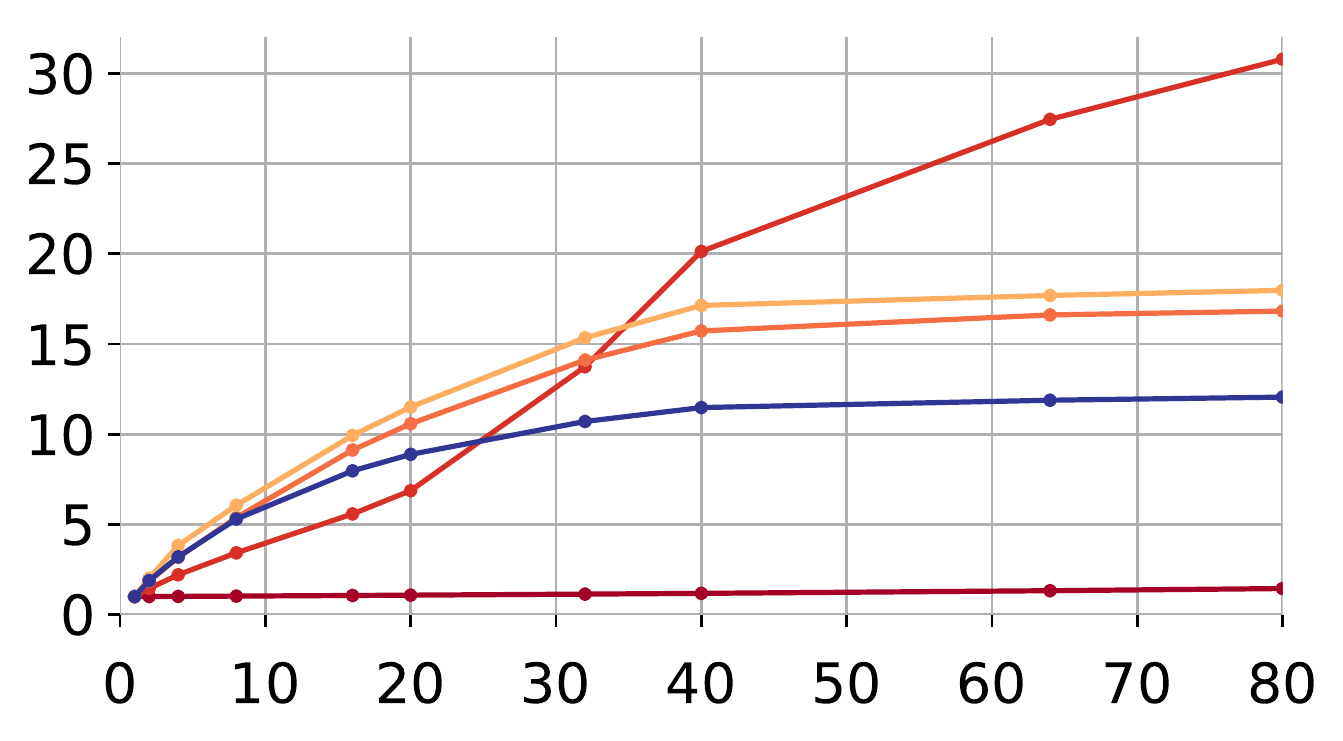}\label{fig:g2_dop_prediction}}
		\subfloat[\gl3]{\includegraphics[width=0.20\textwidth]{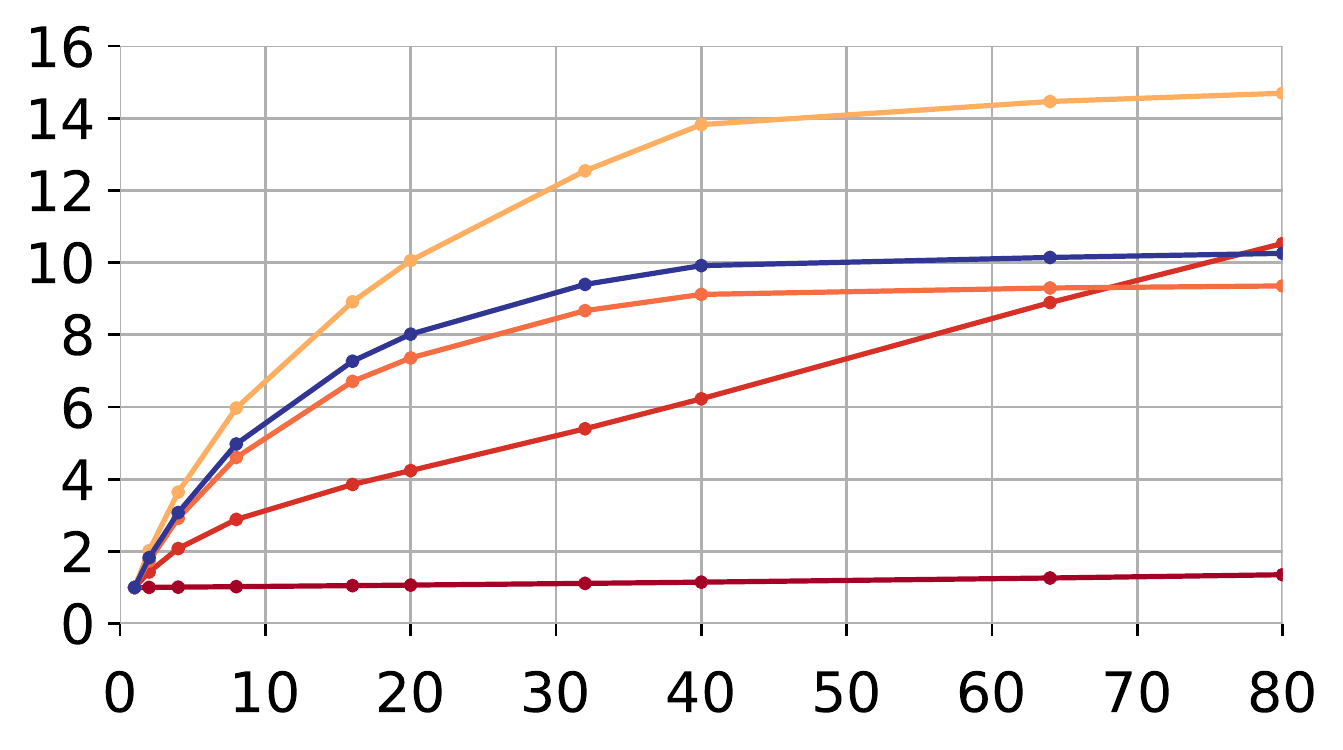}\label{fig:g3_dop_prediction}}
		\subfloat[\gl 4]{\includegraphics[width=0.20\textwidth]{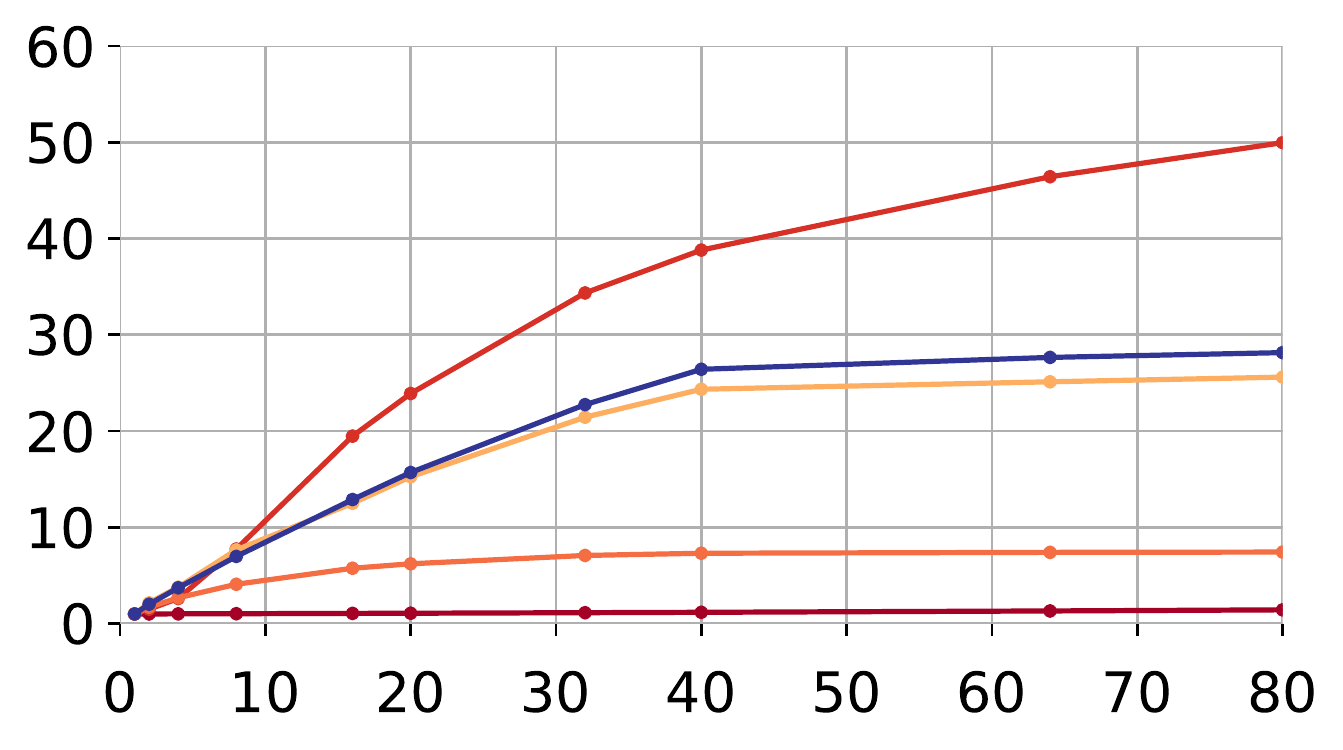}\label{fig:g4_dop_prediction}}
	\caption{Predicted Performance at Per-Query Optimal-DOP --- Query Throughput over DOP 64 (Y-Axis) vs. DOP (X-Axis)} 
	\label{fig:perquery_dop_performance_curve_predict}
%
		\subfloat[\gl 1]{\includegraphics[width=0.20\textwidth]{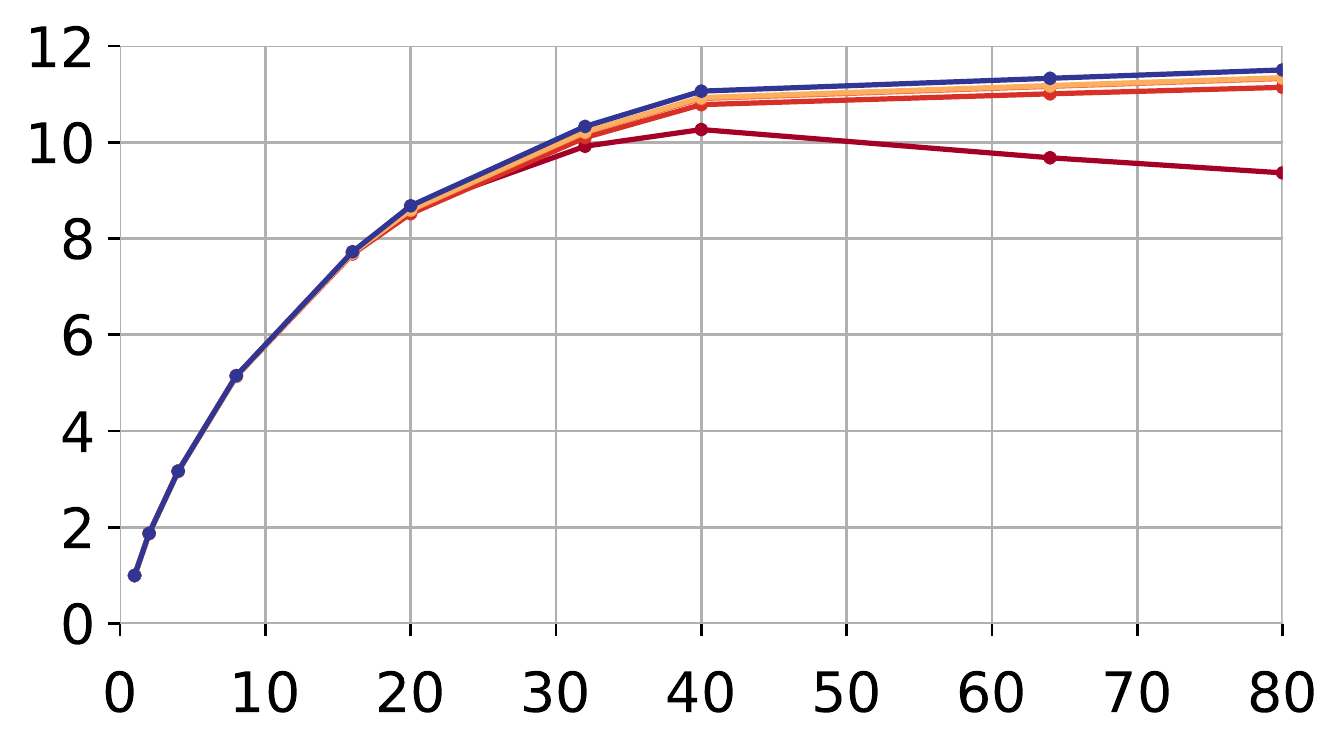}\label{fig:g1_dop_prediction_actual}}
		\subfloat[\gl 2]{\includegraphics[width=0.20\textwidth]{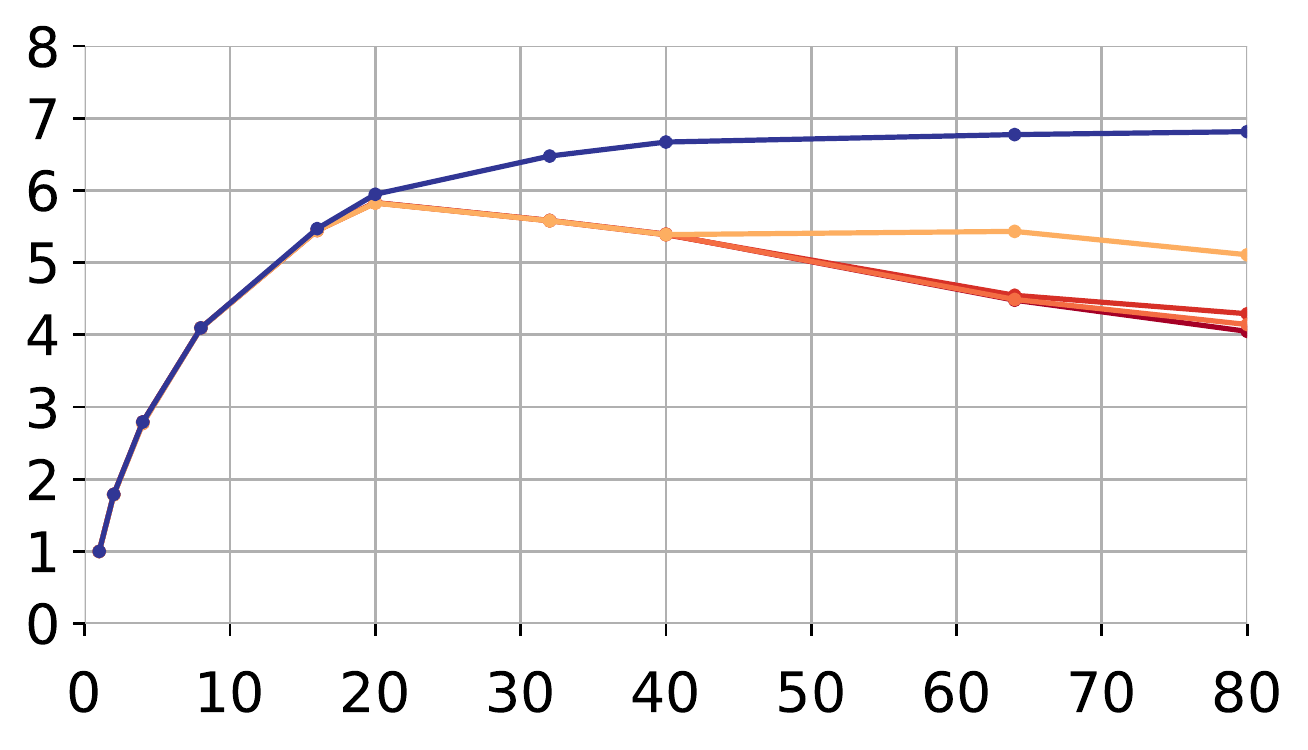}\label{fig:g2_dop_prediction_actual}}
		\subfloat[\gl 2M]{\includegraphics[width=0.20\textwidth]{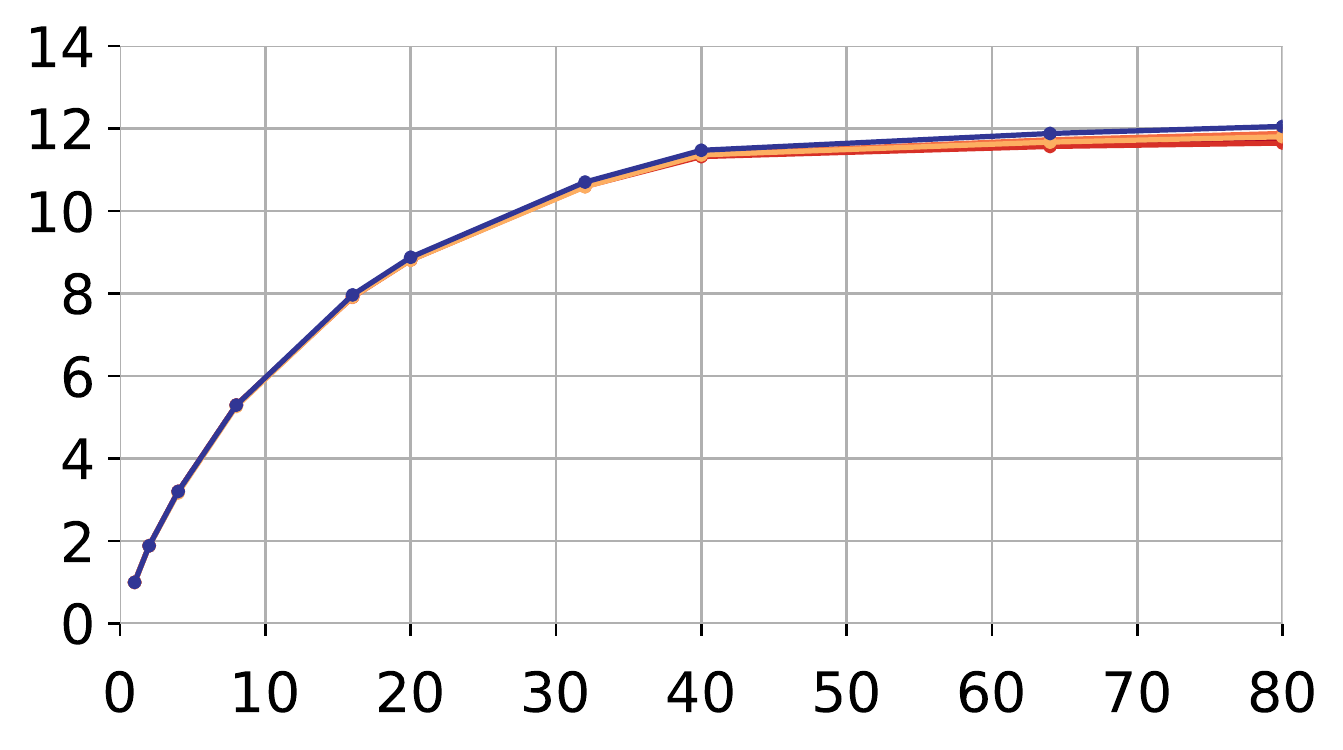}\label{fig:g2_dop_prediction_actual}}
		\subfloat[\gl 3]{\includegraphics[width=0.20\textwidth]{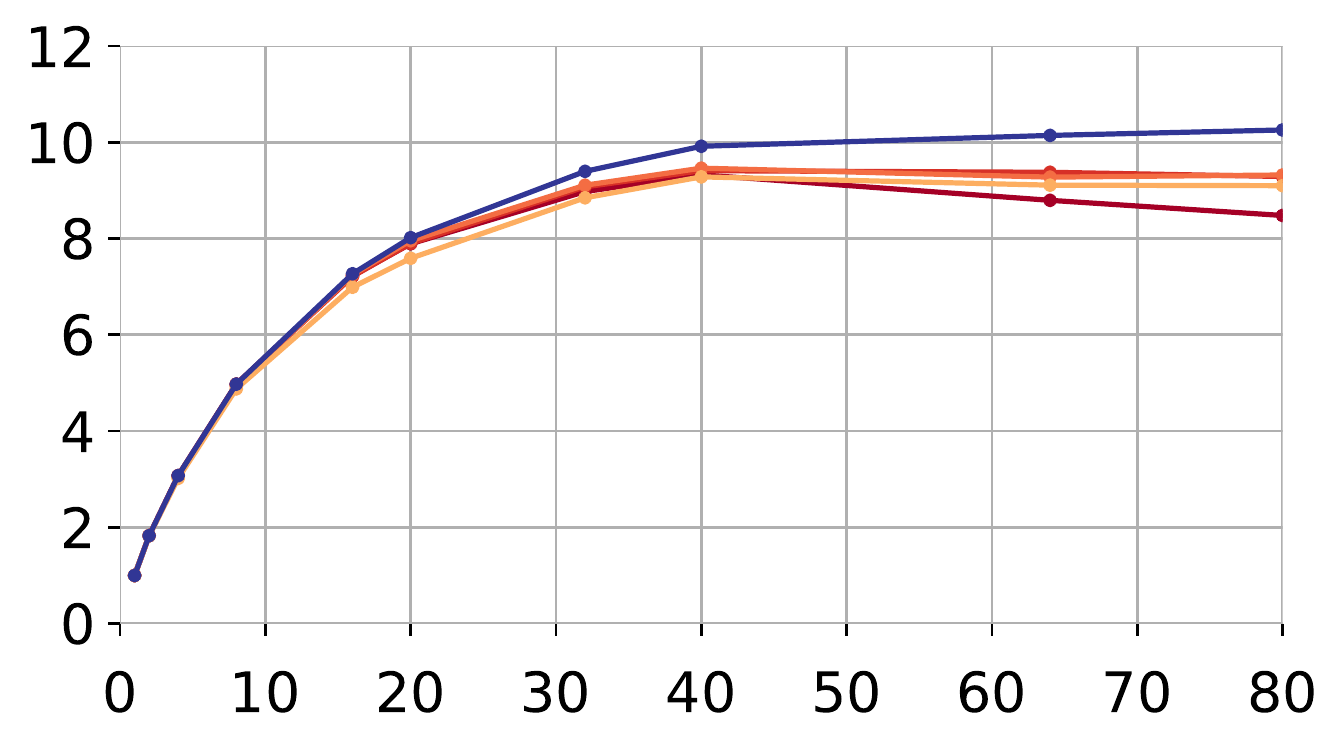}\label{fig:g3_dop_prediction_actual}}
		\subfloat[\gl 4]{\includegraphics[width=0.20\textwidth]{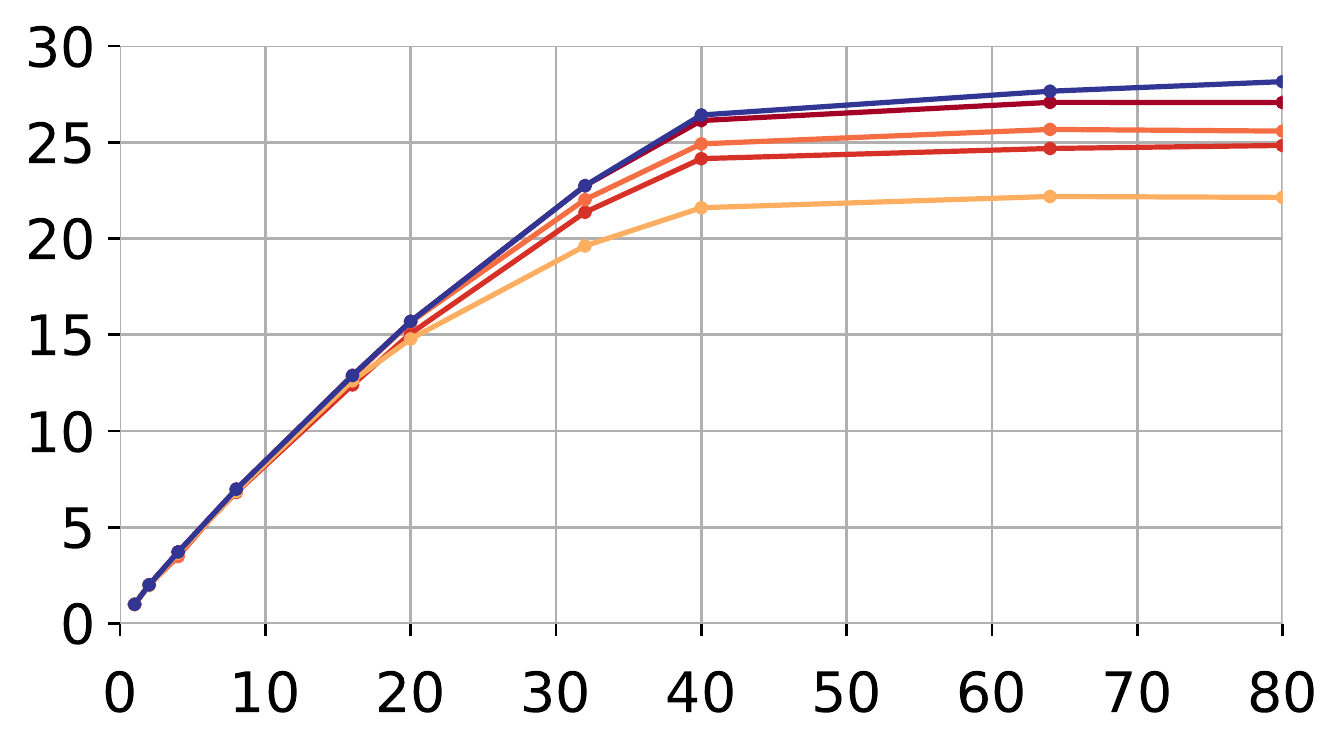}\label{fig:g4_dop_prediction_actual}}
	\caption{Performance at Predicted Per-Query Optimal-DOP --- Query Throughput over DOP 64 (Y-Axis) vs. DOP (X-Axis)}
	\label{fig:perquery_dop_performance_curve_actual}
%
		\subfloat[\gl 1]{\includegraphics[width=0.20\textwidth]{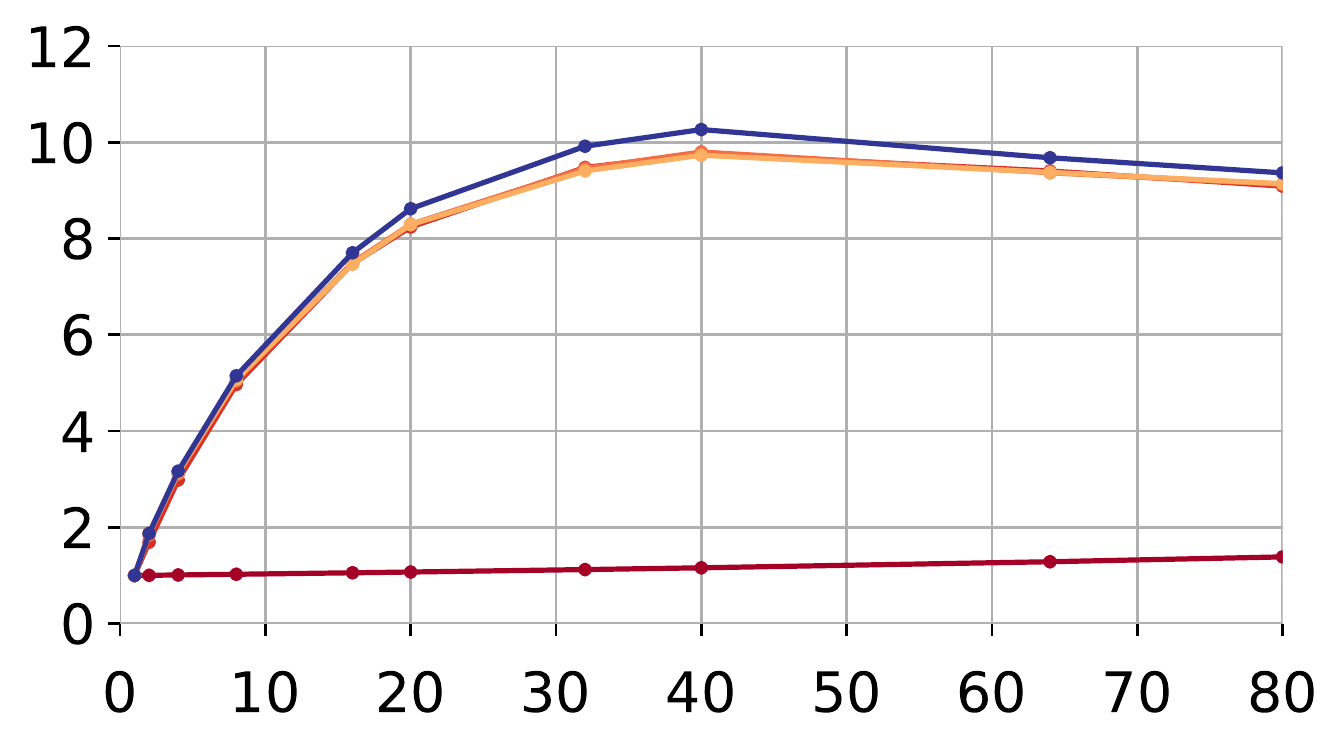}\label{fig:g1_workload_dop_prediction}}
		\subfloat[\gl 2]{\includegraphics[width=0.20\textwidth]{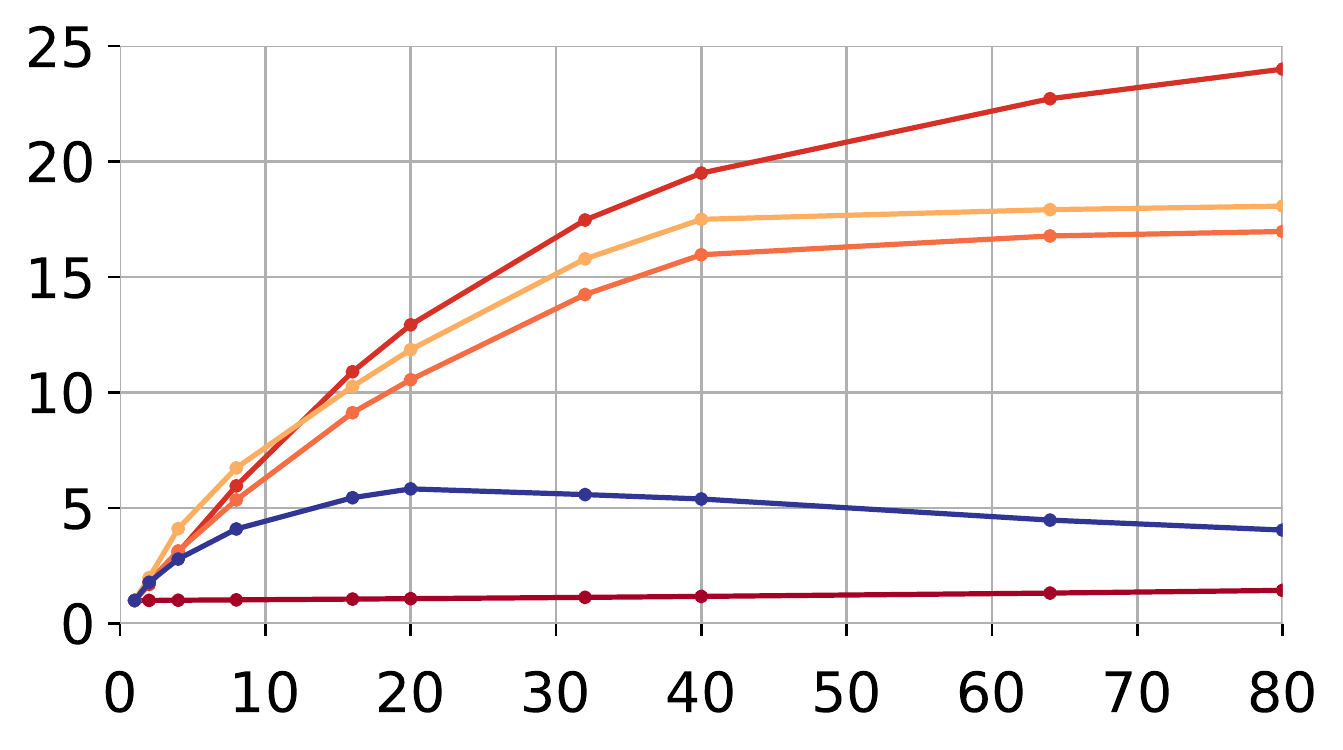}\label{fig:g2_workload_dop_prediction}}
		\subfloat[\gl 2M]{\includegraphics[width=0.20\textwidth]{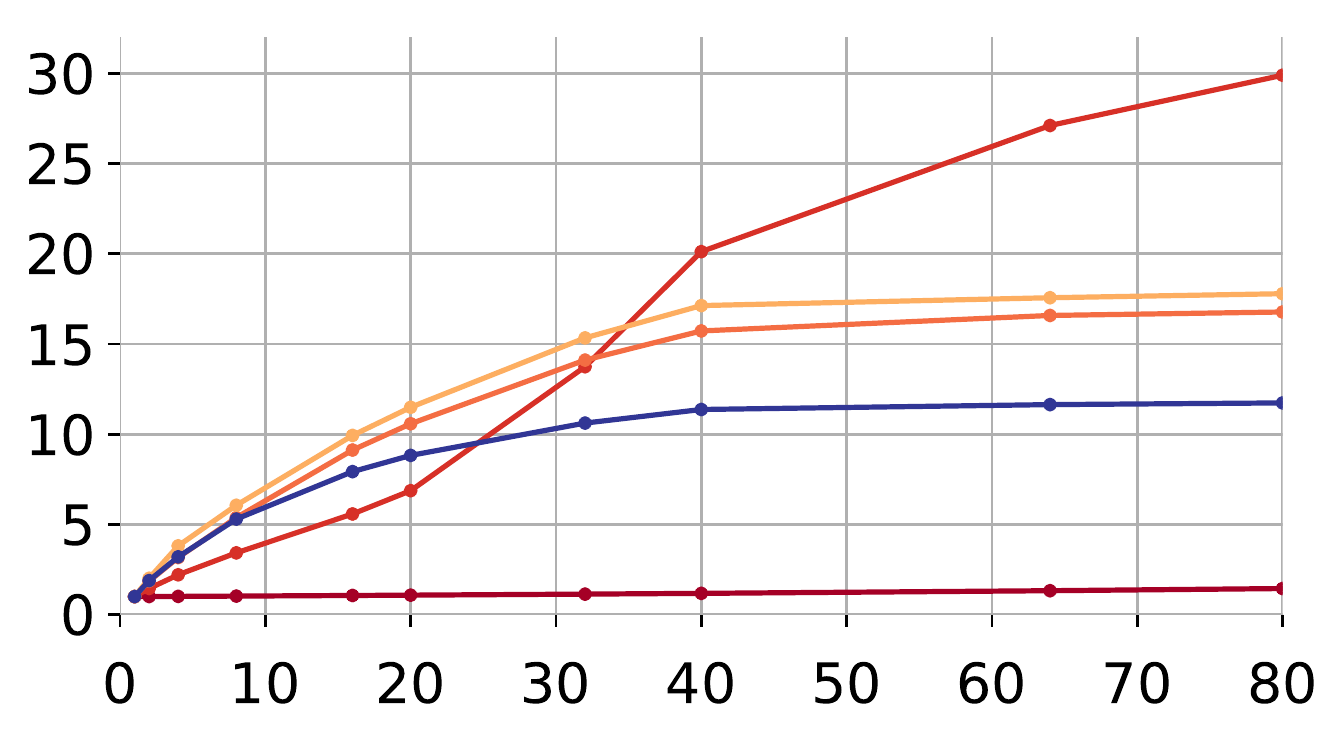}\label{fig:g2m_workload_dop_prediction}}
		\subfloat[\gl 3]{\includegraphics[width=0.20\textwidth]{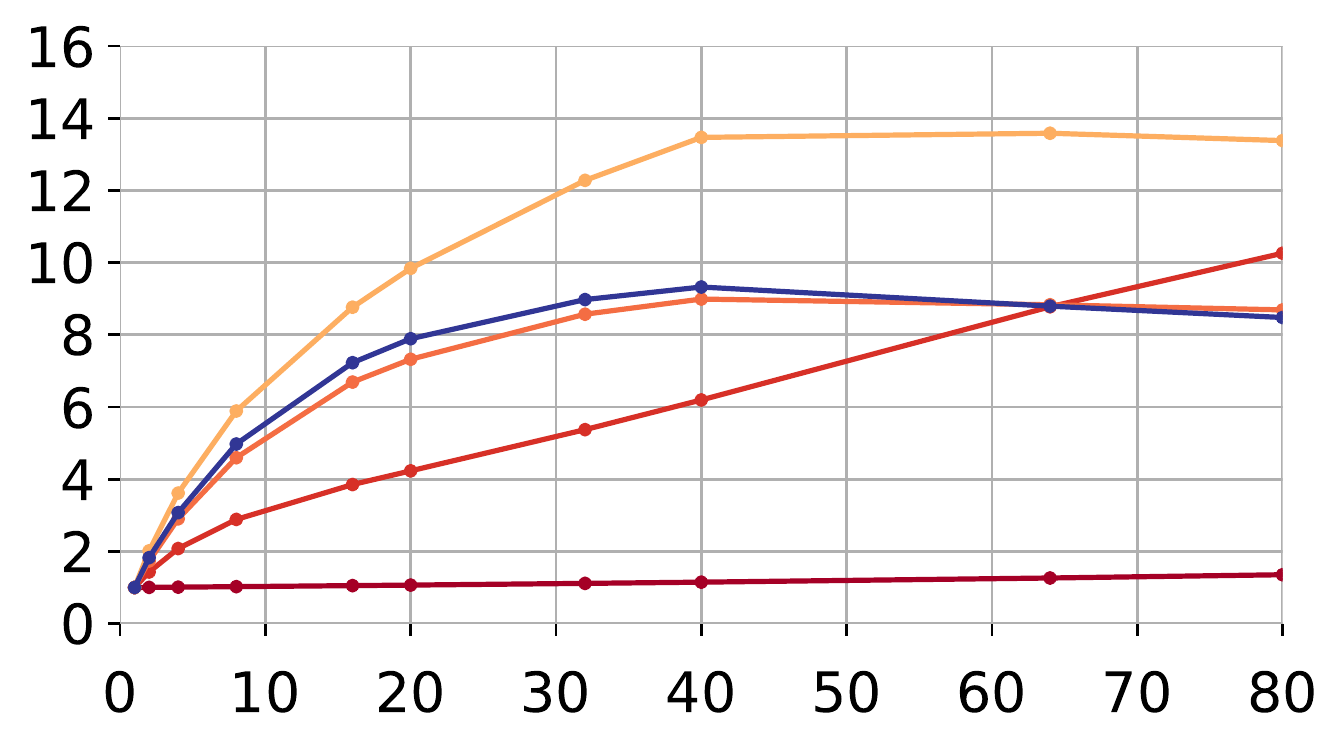}\label{fig:g3_workload_dop_prediction}}
		\subfloat[\gl 4]{\includegraphics[width=0.20\textwidth]{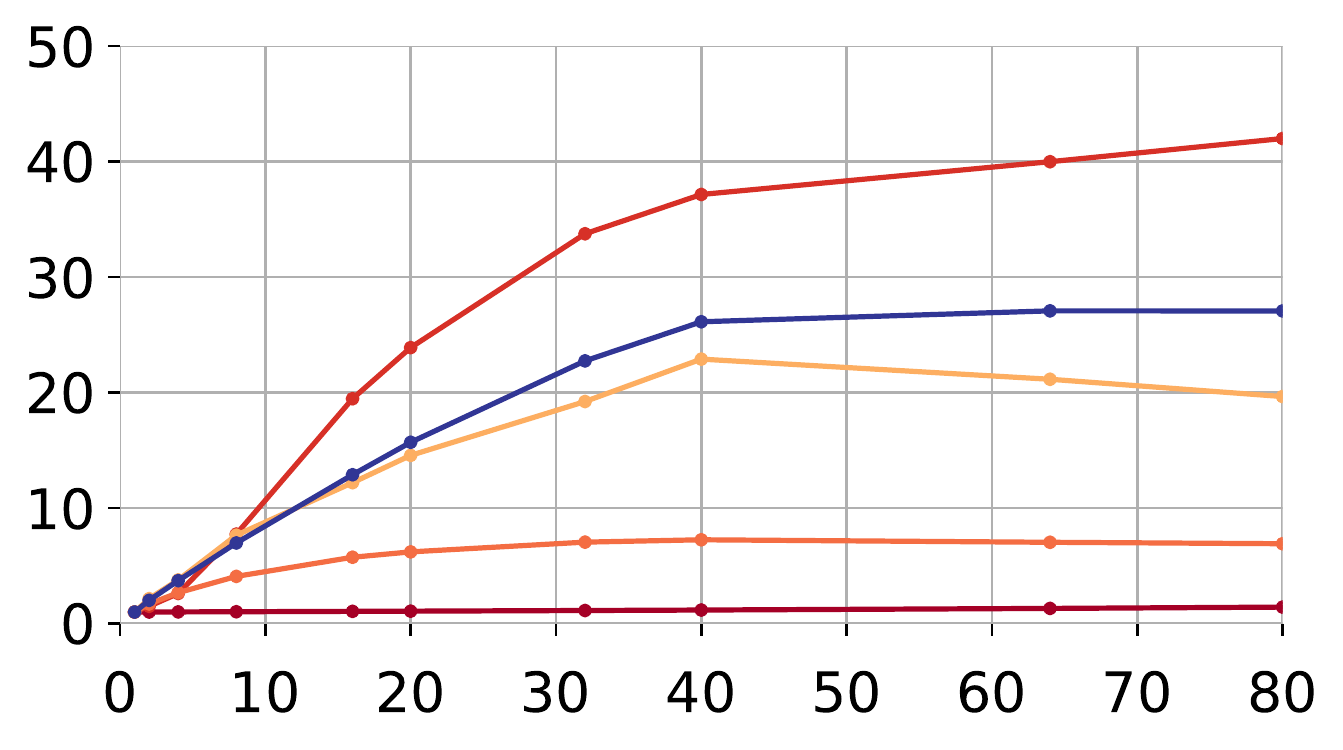}\label{fig:g4_workload_dop_prediction}}
	\caption{Predicted Performance at Workload-Level DOP --- Query Throughput over DOP 64 (Y-Axis) vs. DOP (X-Axis)}
	\label{fig:performance_curve_workload}
\end{figure*}

\subsubsection{Task-Specific Metrics}

\introparagraph{DOP Selection at Individual Query Level}
We now consider the task defined in Section~\ref{sec:task_desc} of selecting the DOP per query Level. We use the metric  $\mathsf{TQ}(W)$ as defined in Table \ref{tab:comparison_metrics}. We compare $\mathsf{TQ}(W)$ given by different models and the query throughputs given by executing queries at DOP 40, 80, the {\em actual optimal DOP  (OPT)} 
for each $P_i \in W$, the {\em workload optimal DOP (WORKLOAD)} for all $P_i \in W$,  with all results being normalized over the throughput at DOP 64 (default value). The comparison results at different generalization levels are shown in Figure \ref{fig:individual_dop_selection}. We first observe that in \gl 1,  all  models except LR lead to query throughput close to  OPT. 

 At \gl 2, we observe that no model is able to predict DOPs that lead to throughput comparable to WORKLOAD in most of the cases. Though XGBoost seems to give performance much closer to WORKLOAD in some cases (Fold 2-5), the results are more likely to be accidental. After investigation, we have found that the root cause behind this  performance gap is the \textit{mismatch} between the data distribution in training and test data: non-trivial disk spilling is observed in queries from the same query template in which $Qd$ (as shown in Figure~\ref{fig:generalization2-examples}) comes from (generated from different random seeds) due to the increase of memory requirements at high DOP values (DOP $>$ 20), which in turn causes a long running time that dominates the whole test workload latency. The observation suggests that a good understanding of the structural differences between the training and test workloads is important before applying ML-based approach. Though many recent works \cite{ding2019ai, marcus2018deep} exploring ML-based techniques in DBMS problems mention the distribution difference between the training  and test data (queries), to the best of our knowledge, we are the first one to explicitly point out such a discrepancy. Meanwhile, the result also reveals that being able predict the memory requirement for a given query with a known query plan, is helpful for DOP tuning. We have verified our analysis by intentionally removing $Qd$-type queries and the results are shown in Figure \ref{fig:gen_2_mem}. Though the performance gain opportunity itself becomes significantly smaller (the performance gap between actual optimal DOP values and the default DOP), in most cases, most models do not select DOPs that result in performance regression and the best-performing model RF often select DOPs giving performance better than WORKLOAD. 

At \gl 3, performance given by models except for LR often matches up OPT or WORKLOAD (Fold 1-4). For Fold 5, since models are trained on plan-dop pairs generated by TPC-DS 100 queries \& database instance and tested on TPC-DS 1000 queries \& database instance (in which disk-spilling is more severe than that being observed in TPC-DS 100), the distribution mismatch (similar to previous reasoning for \gl 2) results in performance regression. 

At \gl 4, the opportunity for throughput improvement over DOP 64 is small (the gap between OPT and throughput at DOP 64 is only $5\% - 10\%$), no model selects DOP values giving performance comparable to OPT, and sometimes models show performance regression compared to the default DOP configuration (64). Interestingly, LR seems to be the best-performing model if solely looking at this task-specific performance metric at \gl 4. The cause of this phenomenon is that TPC-H queries, from which the test data is generated, scale very well with increasing DOP on SQL Server. And LR {\em always} chooses DOP 80 due to its linearity. At first it may seem like the poor performance shown in other models is due to the fact that the degree of commonality between training and testing queries in \gl4 is low, however,
it is noticeable that there is no distinguishable difference between the SPE distributions of \gl3 and \gl4 (Figure \ref{fig:gen3_spe} \ref{fig:gen4_spe}), while RF does a fairly good job in DOP selection for \gl3. 
After careful investigation, we find out that for most of the queries, RF and XGBoost are able to capture the speedup trend accurately, and the relatively poor performance seems more likely to be caused by the specific characteristics of TPC-H queries (testing queries in \gl4) - they scale really well with increasing DOP (see Figure~\ref{fig:g4_dop_prediction_actual}), suggesting large performance gap between query executions at different DOP values and small latency at high DOP (e.g., nearly 30X speedup at DOP 80 compared to performance at DOP 1). Keeping this fact in mind and considering the DS1000/H1000 training/testing split (Fold 2), RF and XGBoost failed to select the near-optimal DOP for two relatively long-running queries in the testing split (RF selects DOP 20 and XGBoost selects DOP 16 while the actual optimal DOPs are 80),  resulting in the performance gaps observed in Figure \ref{fig:individual_dop_selection}. 

\subsubsection{Cost-Performance Trade-off DOP Selection}

While making the optimal choice of DOP configuration at individual query is useful, selecting the optimal DOP at \textit{workload-level}, although is sub-optimal, could ease DOP configuration. Figure \ref{fig:performance_curve_workload} visualizes the actual speedup curve and the predicted speedup curves given by different models with increasing degree of parallelism. The speedup curve can be used for selecting DOP for a given set of queries based on a resource budget, or by simply choosing the optimal point without any constraints.
For example, looking at Figure \ref{fig:g1_workload_dop_prediction}, the speedup by increasing DOP from $32$ to $40$ is insignificant but the hardware provision cost could increase by $25\%$ (assuming the hardware provision cost is proportion to the number of logical cores). 

We observe that the best-performing model RF is able to approximately capture the trend of the actual speedup curves in \gl 1, 2M and 3 considering both per-query optimal-DOP (Figure \ref{fig:perquery_dop_performance_curve_predict}) and workload-level DOP (Figure \ref{fig:performance_curve_workload}). Besides, 
RF is also able to select per-query level DOPs for a given workload, leading to performance that is close to the performance given by running each query at the actual optimal-DOP given the specified maximum DOP possible (Figure \ref{fig:perquery_dop_performance_curve_actual}). However, there is a large gap between the actual speedup curve and the curve predicted by RF in \gl 2 and \gl4, where XGBoost is performing better in predicting the speedup curve in \gl4 (Figure~\ref{fig:g4_dop_prediction},\ref{fig:g4_workload_dop_prediction} )-- in contrast to our observations for task-agnostic metrics.
This gap is caused by the heterogeneous predictions on different plan-DOP pairs in the test data: the prediction output of a few plan-DOP pairs dominates the predicted performance, and hence the actual performance of other plan-DOP pairs is not accurately reflected. On the other hand, XGBoost fails to capture the performance difference (e.g., query performance improves) for a few plan-DOP pairs that are important for DOP selection (Figure~\ref{fig:g4_dop_prediction_actual}), which though are not critical in 
predicting the overall speedup trend (Figure~\ref{fig:g4_dop_prediction},\ref{fig:g4_workload_dop_prediction}). Based on this observation, we argue that merely looking at the task-agnostic metrics might not be always the best way to evaluate the performance of the model, but rather the task-specific ones should be considered together. For example, RF is doing well in terms of per-query DOP selection in \gl 4 while XGBoost is preferred when predicting the view of cost-performance trade-offs is important. Meanwhile, a better learning objective, that forces the model to more accurately learn the absolute and relative difference between the query latency at different DOPs  is critical to make ML-based techniques for query DOP tuning more feasible. 

 

\begin{table}[!t]
\caption{RF Distribution of Relative Prediction Error (RPE) of different featurization (\gl 1)}
\begin{center}
\setlength{\tabcolsep}{1pt}
\begin{tabular}{ c c c c c c }
\toprule
\textbf{RPE} & $F$ & $F\backslash \{\fcard\}$ & $F\backslash \{\fcost\}$ & $F\backslash \{\fcount\}$ & $F\backslash \{\fweight\}$ \\
$\leq{0.1}$ & \textbf{78.8\%}  & 77.7\%  & 77.0\%  & 77.4\%  & 77.8\% \\
$\leq{0.2}$ & \textbf{88.4\%}  & 88.0\%  & 87.4\%  & 87.4\%  & 88.2\% \\
$\leq{0.3}$ & \textbf{91.8\%}  & \textbf{91.8\%}  & 91.6\%  & 91.1\%  & 91.6\% \\
$\leq{0.4}$ & 92.9\%  & 93.2\%  & 93.1\%  & 92.8\%  & \textbf{93.3\%} \\
$\leq{0.5}$ & 94.2\%  & 94.2\%  & 94.1\%  & \textbf{94.4\%}  & 94.3\% \\
$\leq{0.6}$ & 94.9\%  & 94.4\%  & 94.8\%  & \textbf{95.2\%}  & 94.9\% \\
$\leq{0.7}$ & 95.4\%  & 95.2\%  & 95.3\%  & \textbf{95.9\%}  & 95.3\% \\
$\leq{0.8}$ & 96.2\%  & 95.9\%  & \textbf{96.4\%}  & 96.3\%  & 96.3\% \\
$\leq{0.9}$ & 96.5\%  & 96.3\%  & 96.5\%  & \textbf{96.8\%}  & 96.5\% \\
$\leq{1.0}$ & 97.3\%  & 97.0\%  & 97.2\%  & \textbf{97.4\%}  & \textbf{97.4\%} \\
\bottomrule
\end{tabular}
\label{tab:featurization_rpe_g1}
%
\vspace{15pt}
\caption{RF Distribution of Speedup Prediction Error (SPE) of different featurization (\gl 1)}
\vspace{5pt}
\setlength{\tabcolsep}{1pt}
\begin{tabular}{ c c c c c c }
\toprule
\textbf{SPE} & $F$ & $F\backslash \{\fcard\}$ & $F\backslash \{\fcost\}$ & $F\backslash \{\fcount\}$ & $F\backslash \{\fweight\}$ \\
$\leq{0.001}$ & 17.1\%  & 17.7\%  & \textbf{19.6\%}  & 17.3\%  & 17.5\% \\
$\leq{0.005}$ & 67.1\%  & 66.8\%  & \textbf{69.4\%}  & 68.3\%  & 67.0\% \\
$\leq{0.010}$ & 79.3\%  & 79.0\%  & \textbf{80.4\%}  & 78.6\%  & 78.8\% \\
$\leq{0.050}$ & \textbf{94.8\%}  & 93.9\%  & 94.3\%  & 94.7\%  & \textbf{94.8\%} \\
$\leq{0.100}$ & 97.4\%  & 97.4\%  & 97.3\%  & \textbf{97.6\%}  & 97.3\% \\
\bottomrule
\end{tabular}
\label{tab:featurization_spe_g1}
\end{center}
\end{table}

\begin{table}[!t]
\caption{RF Distribution of Relative Prediction Error (RPE) of different featurization (\gl 2)}
\begin{center}
\setlength{\tabcolsep}{1pt}
\begin{tabular}{ c c c c c c }
\toprule
\textbf{RPE} & $F$ & $F\backslash \{\fcard\}$ & $F\backslash \{\fcost\}$ & $F\backslash \{\fcount\}$ & $F\backslash \{\fweight\}$ \\
$\leq{0.1}$ & \textbf{9.0\%}  & 8.2\%  & 7.3\%  & 7.6\%  & 7.3\% \\
$\leq{0.2}$ & 14.5\%  & 15.0\%  & \textbf{15.1\%}  & 13.2\%  & 14.4\% \\
$\leq{0.3}$ & 22.2\%  & 21.9\%  & 21.3\%  & \textbf{23.6\%}  & 23.5\% \\
$\leq{0.4}$ & \textbf{32.4\%}  & 29.4\%  & 26.6\%  & 31.4\%  & 32.2\% \\
$\leq{0.5}$ & \textbf{39.7\%}  & 36.9\%  & 36.4\%  & 36.4\%  & 39.3\% \\
$\leq{0.6}$ & 42.6\%  & 40.7\%  & \textbf{44.1\%}  & 40.2\%  & 41.7\% \\
$\leq{0.7}$ & 46.0\%  & 46.5\%  & \textbf{48.2\%}  & 42.4\%  & 46.2\% \\
$\leq{0.8}$ & 49.1\%  & 51.1\%  & \textbf{51.7\%}  & 46.5\%  & 49.6\% \\
$\leq{0.9}$ & 52.2\%  & \textbf{57.5\%}  & 56.2\%  & 51.5\%  & 52.2\% \\
$\leq{1.0}$ & 57.1\%  & \textbf{62.2\%}  & 60.7\%  & 58.0\%  & 57.4\% \\
\bottomrule
\end{tabular}
\label{tab:featurization_rpe_g2}
%
\vspace{15pt}
\caption{RF Distribution of Speedup Prediction Error (SPE) of different featurization (\gl 2)}
\vspace{5pt}
\setlength{\tabcolsep}{1pt}
\begin{tabular}{ c c c c c c }
\toprule
\textbf{SPE} & $F$ & $F\backslash \{\fcard\}$ & $F\backslash \{\fcost\}$ & $F\backslash \{\fcount\}$ & $F\backslash \{\fweight\}$ \\
$\leq{0.001}$ & 1.1\%  & 0.8\%  & \textbf{1.4\%}  & 1.1\%  & 1.1\% \\
$\leq{0.005}$ & 11.2\%  & 10.1\%  & 10.1\%  & 7.2\%  & \textbf{12.1\%} \\
$\leq{0.010}$ & 27.9\%  & 21.7\%  & \textbf{28.9\%}  & 18.5\%  & 25.1\% \\
$\leq{0.050}$ & 76.7\%  & 74.9\%  & 71.4\%  & 79.9\%  & \textbf{81.8\%} \\
$\leq{0.100}$ & 82.6\%  & 83.9\%  & 84.0\%  & \textbf{85.9\%}  & 84.2\% \\
\bottomrule
\end{tabular}
\label{tab:featurization_spe_g2}
\end{center}
\end{table}

\begin{table}[!t]
\caption{RF Distribution of Relative Prediction Error (RPE) of different featurization (\gl 3)}
\begin{center}
\setlength{\tabcolsep}{1pt}
\begin{tabular}{ c c c c c c }
\toprule
\textbf{RPE} & $F$ & $F\backslash \{\fcard\}$ & $F\backslash \{\fcost\}$ & $F\backslash \{\fcount\}$ & $F\backslash \{\fweight\}$ \\
$\leq{0.1}$ & 0.7\%  & 0.4\%  & 0.5\%  & 0.8\%  & \textbf{0.8\%} \\
$\leq{0.2}$ & 5.7\%  & 4.5\%  & 5.0\%  & \textbf{6.3\%}  & 5.4\% \\
$\leq{0.3}$ & 12.7\%  & 10.4\%  & 10.4\%  & 12.0\%  & \textbf{12.7\%} \\
$\leq{0.4}$ & \textbf{19.8\%}  & 16.7\%  & 18.6\%  & 18.3\%  & 19.7\% \\
$\leq{0.5}$ & \textbf{27.5\%}  & 24.0\%  & 24.9\%  & 25.1\%  & 27.3\% \\
$\leq{0.6}$ & \textbf{33.6\%}  & 32.7\%  & 33.5\%  & 33.0\%  & 33.3\% \\
$\leq{0.7}$ & 38.8\%  & \textbf{39.4\%}  & 38.5\%  & 38.1\%  & 38.9\% \\
$\leq{0.8}$ & 42.4\%  & \textbf{44.4\%}  & 43.7\%  & 42.3\%  & 42.5\% \\
$\leq{0.9}$ & 46.1\%  & 48.7\%  & \textbf{48.8\%}  & 46.0\%  & 45.9\% \\
$\leq{1.0}$ & 49.1\%  & 51.2\%  & \textbf{51.9\%}  & 48.8\%  & 50.0\% \\

\bottomrule
\end{tabular}
\label{tab:featurization_rpe_g3}
%
\vspace{15pt}
\caption{RF Distribution of Speedup Prediction Error (SPE) of different featurization (\gl 3)}
\vspace{5pt}
\setlength{\tabcolsep}{1pt}
\begin{tabular}{ c c c c c c }
\toprule
\textbf{SPE} & $F$ & $F\backslash \{\fcard\}$ & $F\backslash \{\fcost\}$ & $F\backslash \{\fcount\}$ & $F\backslash \{\fweight\}$ \\
$\leq{0.001}$ & \textbf{0.0\%}  & \textbf{0.0\%}  & \textbf{0.0\%}  & \textbf{0.0\%}  & \textbf{0.0\%} \\
$\leq{0.005}$ & 1.4\%  & 0.9\%  & \textbf{1.9\%}  & 0.4\%  & 1.7\% \\
$\leq{0.010}$ & 7.6\%  & 7.8\%  & \textbf{10.9\%}  & 5.5\%  & 7.0\% \\
$\leq{0.050}$ & 56.0\%  & 63.4\%  & \textbf{63.6\%}  & 58.3\%  & 57.7\% \\
$\leq{0.100}$ & 76.9\%  & \textbf{80.8\%}  & 80.5\%  & 77.4\%  & 79.0\% \\

\bottomrule
\end{tabular}
\label{tab:featurization_spe_g3}
\end{center}
\end{table}

\begin{table}[!t]
\caption{RF Distribution of Relative Prediction Error (RPE) of different featurization (\gl 4)}
\begin{center}
\setlength{\tabcolsep}{1pt}
\begin{tabular}{ c c c c c c }
\toprule
\textbf{RPE} & $F$ & $F\backslash \{\fcard\}$ & $F\backslash \{\fcost\}$ & $F\backslash \{\fcount\}$ & $F\backslash \{\fweight\}$ \\
$\leq{0.1}$ & 1.5\%  & \textbf{2.3\%}  & 0.0\%  & 0.0\%  & 1.5\% \\
$\leq{0.2}$ & 7.0\%  & \textbf{10.9\%}  & 10.5\%  & 5.3\%  & 6.3\% \\
$\leq{0.3}$ & 16.9\%  & 19.9\%  & \textbf{20.7\%}  & 15.7\%  & 19.7\% \\
$\leq{0.4}$ & 26.1\%  & 25.1\%  & \textbf{27.4\%}  & 23.4\%  & 27.4\% \\
$\leq{0.5}$ & 33.4\%  & 33.1\%  & \textbf{36.7\%}  & 31.0\%  & 32.2\% \\
$\leq{0.6}$ & 38.8\%  & 40.1\%  & \textbf{46.7\%}  & 36.3\%  & 41.4\% \\
$\leq{0.7}$ & 48.8\%  & 46.5\%  & \textbf{53.3\%}  & 38.5\%  & 49.2\% \\
$\leq{0.8}$ & 52.8\%  & 56.8\%  & \textbf{57.2\%}  & 41.0\%  & 53.0\% \\
$\leq{0.9}$ & 58.0\%  & \textbf{66.9\%}  & 62.1\%  & 45.9\%  & 58.6\% \\
$\leq{1.0}$ & 62.3\%  & \textbf{71.3\%}  & 63.3\%  & 47.1\%  & 61.3\% \\
\bottomrule
\end{tabular}
\label{tab:featurization_rpe_g4}
%
\vspace{15pt}
\caption{RF Distribution of Speedup Prediction Error (SPE) of different featurization (\gl 4)}
\vspace{5pt}
\setlength{\tabcolsep}{1pt}
\begin{tabular}{ c c c c c c }
\toprule
\textbf{SPE} & $F$ & $F\backslash \{\fcard\}$ & $F\backslash \{\fcost\}$ & $F\backslash \{\fcount\}$ & $F\backslash \{\fweight\}$ \\
$\leq{0.001}$ & \textbf{0.0\%}  & \textbf{0.0\%}  & \textbf{0.0\%}  & \textbf{0.0\%}  & \textbf{0.0\%} \\
$\leq{0.005}$ & 1.4\%  & 0.9\%  & \textbf{1.9\%}  & 0.4\%  & 1.7\% \\
$\leq{0.010}$ & 7.6\%  & 7.8\%  & \textbf{10.9\%}  & 5.5\%  & 7.0\% \\
$\leq{0.050}$ & 56.0\%  & 63.4\%  & \textbf{63.6\%}  & 58.3\%  & 57.7\% \\
$\leq{0.100}$ & 76.9\%  & \textbf{80.8\%}  & 80.5\%  & 77.4\%  & 79.0\% \\
\bottomrule
\end{tabular}
\label{tab:featurization_spe_g4}
\end{center}
\end{table}

\section{Related Work}
\label{sec:related}

Query Performance Prediction (QPP) is a well-studied problem~\cite{akdere2012learning,ganapathi2009predicting,li2012robust,wu:qpp-concurrent-queries:vldb:2013,marcus2019plan}, but not in a setting with intra-query parallelism. Early work in DOP management~\cite{mehta1995managing} studied the exploitation of intra-operator parallelism in a multi-query environment for shared-nothing parallel database systems using a simplified simulation model. More recent works have developed analytical models for query parallelism in BigData execution frameworks~\cite{Rajan:perforator:socc:2016}.
To the best of our knowledge, this is the first paper to do a comparative study of ML techniques for QPP for multithreaded query execution and evaluate query DOP tuning for a commercial-grade RDBMS on a modern multicore server.


\introparagraph{ML for Query Performance Prediction}
Prior work has explored the use of per-operator models~\cite{akdere2012learning,li2012robust,marcus2019plan} for QPP. The per-operator latency estimates can be added together~\cite{akdere2012learning,li2012robust} or combined usings plan/subplan-level estimates~\cite{akdere2012learning} or DNNs~\cite{marcus2019plan} to get query-level estimates. While operator-level modeling may have the potential for better generalization to unknown plans, there are several challenges in using this approach for highly-optimized RDBMSs.

\begin{figure}[h]
\vspace{-10pt}
		\subfloat{\includegraphics[width=0.24\textwidth]{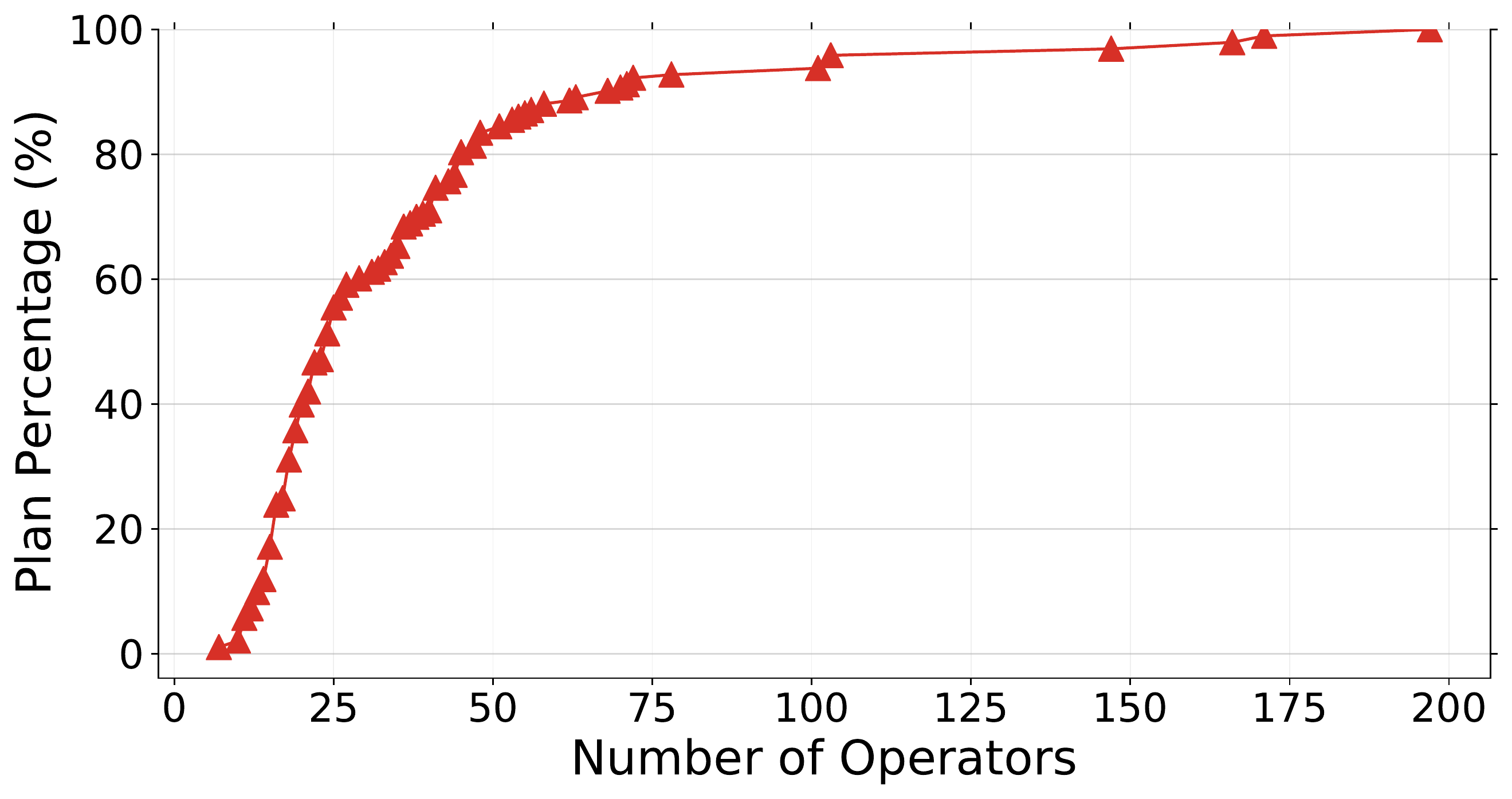}}
		\subfloat{\includegraphics[width=0.24\textwidth]{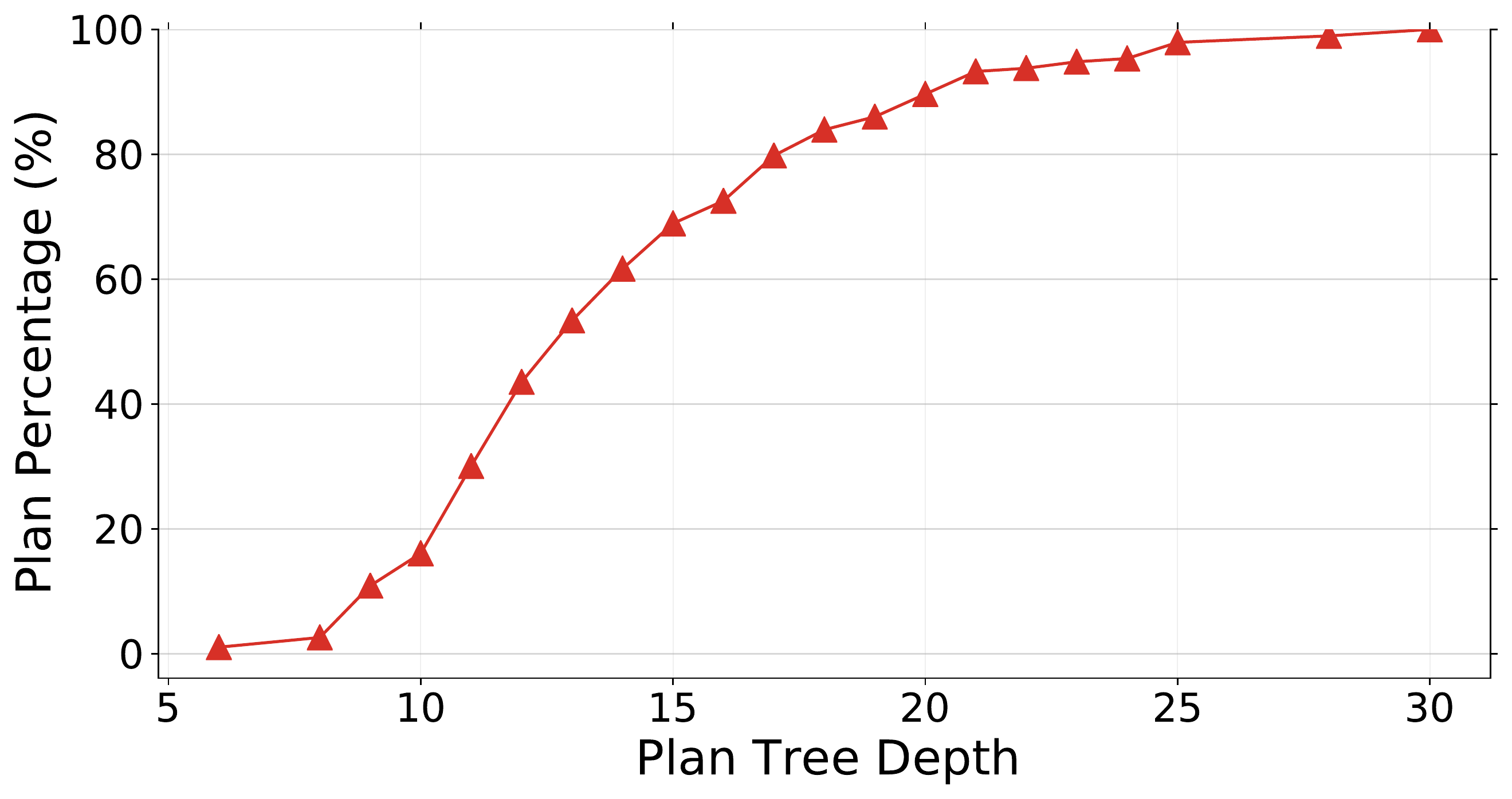}}\\
	\caption{TPC-DS1000 plan size \& depth distribution.}
	\label{fig:operator-distrib}
\end{figure}

First, query estimates based on per-operator modeling can incur significant complexity and cost for plans with a large number of operators.
Figure~\ref{fig:operator-distrib} shows the distributions of the number of operators (plan size) and height of the relational operator tree (plan depth) in the plans for TPC-DS 1000 queries that we study. About 6.2\% of plans have size of more than 100, with a maximum of 197, and about 10\% have a depth of 20 or more. Large plans have high training costs while deep plans increase difficulty of training QPPNet-like DNN-based models, critical path lengths and inference overhead at query execution time.

Second, building a query-level combined model from per-operator models is not always feasible when there is pipelined parallel execution, because of the temporal overlap of and contention for shared resources by inter- and intra-operator threads. The semantics of isolated per-operator latencies are not well-defined in such a setting, and it may prevents from extracting and composing operator-level estimates to derive query-level estimates as Figure \ref{fig:qpp_net_comparison} suggests.

\introparagraph{Automated Database Tuning}
Prior work has also explored techniques for automatically tuning database systems to improve query performance~\cite{Chaudhuri:self-tuning-db:vldb:2007,Duan:ituned:vldb:2009,pavlo2017self,VanAken:ottertune:sgmod:2017}. Our work on DOP tuning can benefit such approaches by providing what-if analyses for different DOP settings and thereby potentially eliminating the need for runtime exploration of this parameter space.

\introparagraph{Dynamic Parallelism}
Recent work on new database systems use dynamic/elastic parallelism where the the parallelism can be increased/decreased at runtime~\cite{Leis:morsel:sigmod:2014,Patel:quickstep:vldb:2018}. Although our ML-based DOP selection approach for performance improvement focuses on a setting with static parallelism, we believe that our approach on providing cost-benefit tradeoff estimates would be useful for resource provisioning in elastic systems as well.



\introparagraph{ML for Query Optimization}
Conventional query optimization in RDBMSs is based on manually constructed cost models and rule-based heuristics. Such strategies usually make certain assumptions about the plan search space (e.g., left-deep tree) and rely heavily on the estimation of statistics required by the cost models such as cardinality estimation-- which can be quite inaccurate, leading to query plans that are far away from optimal. In addition, query optimizers built this way never learn from past experience, and the same query plan with poor performance can be repeatedly generated~\cite{marcus2018towards}. Recent research efforts attempt to enhance and even replace the core optimizer components seen in most of RDBMSs today using \textit{deep learning} (DL) techniques. For example, \cite{marcus2018deep,krishnan2018learning} exploit \textit{deep reinforcement learning} (DRL) to optimize join queries, while \cite{kipf2018learned,ortiz2018learning} propose to use DRL to improve the cardinality estimation. Neo, a query optimizer built based on a set of DL-based models~\cite{marcus2019neo}, has been shown to even outperform state-of-the-art commercial optimizers in some cases. Compared to this line of research, our work focuses on optimizing the query performance \textit{outside of the RDBMS} (an example of resource tuning) rather than touching the optimizer internals, similar in scope to~\cite{pavlo2017self}.

\section{Discussion}
\revise{
Our comparative exploration of ML techniques for tuning DOP in SQL Server indicates that performance gains are possible by using simple tree-ensemble models along with query plan featurization. The performance profiling for queries showing different behaviors at different DOP values suggests that DOP tuning is an important while challenging problem. We have also identified a set of important issues that raise concerns when applying the ML-based techniques for DOP tuning in practice, along with the possible improvements to this work:

\begin{itemize}
\item  Memory consideration: in our present study, we run queries in an environment in which the available memory is {\em fixed} (i.e., static). The utility of ML models learned without considering memory information is limited, preventing their use in new environments with different memory requirements (e.g., machines with different memory sizes, or concurrent query execution in which each stream is under strict memory constraints). We plan to address this deficiency as part of improvements to this work.
\item Concurrent query execution: we plan to utilize the single-query based ML models for DOP tuning in a concurrent environment by an analytical approach, expanding the applicability of our study.
\item Model interpretability: the results given by the black-box ML algorithm should be interpretable in a way similar to that of the rule-based algorithm. The DBAs should be able to see the {\em decision path} suggested by the algorithm in order to decide whether to accept/reject the model output. Meanwhile, interpretability could also help with possible implementation issues inside the targeted RDBMS (e.g., the poor parallelism could be caused by the behavior of operators of certain types).
\item Studying the effects of hyper-threading: since enabling or disabling hyper-threading will affect query performance characteristics, it is important to study such differences and maybe even encode this piece of information into ML models.
\end{itemize} 
}

\section{Conclusion }
We studied the problem of tuning the degree of parallelism via statistical machine learning. We focus our evaluation on Microsoft SQL Server, a popular commercial RDBMS that allows an individual query to execute on multiple cores. 
 In our study, we cast the problem of DOP tuning as a {\em regression} task, and examine how several popular ML models can help with query performance prediction in a multi-core setting. 
 We performed an extensive experimental study comparing these models against a list of performance metrics, and tested how well they generalize in different settings: $(i)$ to queries from the same template, $(ii)$ to queries from a new template, $(iii)$ to instances of different scale, and $(iv)$ to different instances and queries.  
 Our experimental results show that a simple featurization of the input query plan that ignores cost model estimations can accurately predict query performance, capture the speedup trend with respect to the available parallelism, as well as help with automatically choosing an optimal per-query DOP.

\section{Acknowledgement}

We thank Alan Halverson for insightful discussions about the DOP problem and about baseline experimental setup during the initial phase of this work. Zhiwei was supported by Microsoft's Gray Systems Lab (GSL) through a summer internship and Microsoft RA-ships for this research work. We thank Carlo Curino and other members of GSL, and members of the SQL Server team for discussions and feedback on this work. The manuscript was improved by detailed and thoughtful comments from  Remzi Arpaci-Dusseau and other, anonymous, reviewers.


\balance
\bibliographystyle{abbrv}
\bibliography{references}

\begin{thebibliography}{10}

\bibitem{akdere2012learning}
M.~Akdere, U.~{\c{C}}etintemel, M.~Riondato, E.~Upfal, and S.~B. Zdonik.
\newblock Learning-based query performance modeling and prediction.
\newblock In {\em 2012 IEEE 28th International Conference on Data Engineering},
  pages 390--401. IEEE, 2012.

\bibitem{Chaudhuri:self-tuning-db:vldb:2007}
S.~Chaudhuri and V.~Narasayya.
\newblock Self-tuning database systems: A decade of progress.
\newblock In {\em Proceedings of the 33rd International Conference on Very
  Large Data Bases}, VLDB '07, pages 3--14. VLDB Endowment, 2007.

\bibitem{chen2016xgboost}
T.~Chen and C.~Guestrin.
\newblock Xgboost: A scalable tree boosting system.
\newblock In {\em Proceedings of the 22nd acm sigkdd international conference
  on knowledge discovery and data mining}, pages 785--794. ACM, 2016.

\bibitem{ding2019ai}
B.~Ding, S.~Das, R.~Marcus, W.~Wu, S.~Chaudhuri, and V.~R. Narasayya.
\newblock Ai meets ai: Leveraging query executions to improve index
  recommendations.
\newblock In {\em Proceedings of the 2019 International Conference on
  Management of Data}, pages 1241--1258. ACM, 2019.

\bibitem{Duan:ituned:vldb:2009}
S.~Duan, V.~Thummala, and S.~Babu.
\newblock Tuning database configuration parameters with ituned.
\newblock {\em Proc. VLDB Endow.}, 2(1):1246--1257, Aug. 2009.

\bibitem{ganapathi2009predicting}
A.~Ganapathi, H.~Kuno, U.~Dayal, J.~L. Wiener, A.~Fox, M.~Jordan, and
  D.~Patterson.
\newblock Predicting multiple metrics for queries: Better decisions enabled by
  machine learning.
\newblock In {\em 2009 IEEE 25th International Conference on Data Engineering},
  pages 592--603. IEEE, 2009.

\bibitem{graefe1990encapsulation}
G.~Graefe.
\newblock Encapsulation of parallelism in the volcano query processing system.
\newblock {\em ACM SIGMOD Record}, 19(2):102--111, 1990.

\bibitem{kim2014convolutional}
Y.~Kim.
\newblock Convolutional neural networks for sentence classification.
\newblock {\em arXiv preprint arXiv:1408.5882}, 2014.

\bibitem{kingma2014adam}
D.~P. Kingma and J.~Ba.
\newblock Adam: A method for stochastic optimization.
\newblock {\em arXiv preprint arXiv:1412.6980}, 2014.

\bibitem{kipf2018learned}
A.~Kipf, T.~Kipf, B.~Radke, V.~Leis, P.~Boncz, and A.~Kemper.
\newblock Learned cardinalities: Estimating correlated joins with deep
  learning.
\newblock {\em arXiv preprint arXiv:1809.00677}, 2018.

\bibitem{krishnan2018learning}
S.~Krishnan, Z.~Yang, K.~Goldberg, J.~Hellerstein, and I.~Stoica.
\newblock Learning to optimize join queries with deep reinforcement learning.
\newblock {\em arXiv preprint arXiv:1808.03196}, 2018.

\bibitem{Leis:morsel:sigmod:2014}
V.~Leis, P.~Boncz, A.~Kemper, and T.~Neumann.
\newblock Morsel-driven parallelism: A numa-aware query evaluation framework
  for the many-core age.
\newblock In {\em Proceedings of the 2014 ACM SIGMOD International Conference
  on Management of Data}, SIGMOD '14, pages 743--754, New York, NY, USA, 2014.
  ACM.

\bibitem{li2012robust}
J.~Li, A.~C. K{\"o}nig, V.~Narasayya, and S.~Chaudhuri.
\newblock Robust estimation of resource consumption for sql queries using
  statistical techniques.
\newblock {\em Proceedings of the VLDB Endowment}, 5(11):1555--1566, 2012.

\bibitem{marcus2019neo}
R.~Marcus, P.~Negi, H.~Mao, C.~Zhang, M.~Alizadeh, T.~Kraska, O.~Papaemmanouil,
  and N.~Tatbul.
\newblock Neo: A learned query optimizer.
\newblock {\em arXiv preprint arXiv:1904.03711}, 2019.

\bibitem{marcus2018deep}
R.~Marcus and O.~Papaemmanouil.
\newblock Deep reinforcement learning for join order enumeration.
\newblock In {\em Proceedings of the First International Workshop on Exploiting
  Artificial Intelligence Techniques for Data Management}, page~3. ACM, 2018.

\bibitem{marcus2018towards}
R.~Marcus and O.~Papaemmanouil.
\newblock Towards a hands-free query optimizer through deep learning.
\newblock {\em arXiv preprint arXiv:1809.10212}, 2018.

\bibitem{marcus2019plan}
R.~Marcus and O.~Papaemmanouil.
\newblock Plan-structured deep neural network models for query performance
  prediction.
\newblock {\em arXiv preprint arXiv:1902.00132}, 2019.

\bibitem{mehta1995managing}
M.~Mehta and D.~J. DeWitt.
\newblock Managing intra-operator parallelism in parallel database systems.
\newblock In {\em Proceedings of the 21th International Conference on Very
  Large Data Bases}, VLDB '95, pages 382--394, San Francisco, CA, USA, 1995.

\bibitem{cci}
Microsoft.
\newblock Columnstore indexes: Overview.
\newblock
  \url{https://docs.microsoft.com/en-us/sql/relational-databases/indexes/columnstore-indexes-overview?view=sql-server-ver15},
  2018.

\bibitem{azuresqldb-paas}
Microsoft.
\newblock {Azure SQL Database - Platform as a Service}.
\newblock
  \url{https://docs.microsoft.com/en-us/azure/sql-database/sql-database-paas},
  2019.

\bibitem{azuresql-serverless-tier}
Microsoft.
\newblock {Azure SQL Database} serverless (preview).
\newblock
  \url{https://docs.microsoft.com/en-us/azure/sql-database/sql-database-serverless},
  2019.

\bibitem{azuresql-vcore-service-tiers}
Microsoft.
\newblock Choose among the {vCore} service tiers and migrate from the {DTU}
  service tiers.
\newblock
  \url{https://docs.microsoft.com/en-us/azure/sql-database/sql-database-service-tiers-vcore},
  2019.

\bibitem{azuresql}
Microsoft.
\newblock Choose the right {SQL Server} option in {Azure}.
\newblock
  \url{https://docs.microsoft.com/en-us/azure/sql-database/sql-database-paas-vs-sql-server-iaas},
  2019.

\bibitem{azuresql-dtu-resource-limits}
Microsoft.
\newblock Resource limits for single databases using the {DTU}-based purchasing
  model.
\newblock
  \url{https://docs.microsoft.com/en-us/azure/sql-database/sql-database-dtu-resource-limits-single-databases},
  2019.

\bibitem{azuresql-vcore-resource-limits}
Microsoft.
\newblock Resource limits for single databases using the {vCore}-based
  purchasing model.
\newblock
  \url{https://docs.microsoft.com/en-us/azure/sql-database/sql-database-vcore-resource-limits-single-databases},
  2019.

\bibitem{azuresql-dtu-service-tiers}
Microsoft.
\newblock Service tiers in the {DTU}-based purchase model.
\newblock
  \url{https://docs.microsoft.com/en-us/azure/sql-database/sql-database-service-tiers-dtu},
  2019.

\bibitem{sqlserver-2019}
Microsoft.
\newblock {SQL Server} 2019.
\newblock \url{https://www.microsoft.com/en-us/sql-server/sql-server-2019},
  2019.

\bibitem{nambiar2006making}
R.~O. Nambiar and M.~Poess.
\newblock The making of tpc-ds.
\newblock In {\em Proceedings of the 32nd international conference on Very
  large data bases}, pages 1049--1058. VLDB Endowment, 2006.

\bibitem{ortiz2018learning}
J.~Ortiz, M.~Balazinska, J.~Gehrke, and S.~S. Keerthi.
\newblock Learning state representations for query optimization with deep
  reinforcement learning.
\newblock {\em arXiv preprint arXiv:1803.08604}, 2018.

\bibitem{Patel:quickstep:vldb:2018}
J.~M. Patel, H.~Deshmukh, J.~Zhu, N.~Potti, Z.~Zhang, M.~Spehlmann,
  H.~Memisoglu, and S.~Saurabh.
\newblock Quickstep: A data platform based on the scaling-up approach.
\newblock {\em Proc. VLDB Endow.}, 11(6):663--676, Feb. 2018.

\bibitem{pavlo2017self}
A.~Pavlo, G.~Angulo, J.~Arulraj, H.~Lin, J.~Lin, L.~Ma, P.~Menon, T.~C. Mowry,
  M.~Perron, I.~Quah, et~al.
\newblock Self-driving database management systems.
\newblock In {\em CIDR}, volume~4, page~1, 2017.

\bibitem{poess2000new}
M.~Poess and C.~Floyd.
\newblock New tpc benchmarks for decision support and web commerce.
\newblock {\em ACM Sigmod Record}, 29(4):64--71, 2000.

\bibitem{Rajan:perforator:socc:2016}
K.~Rajan, D.~Kakadia, C.~Curino, and S.~Krishnan.
\newblock Perforator: Eloquent performance models for resource optimization.
\newblock In {\em Proceedings of the Seventh ACM Symposium on Cloud Computing},
  SoCC '16, pages 415--427, New York, NY, USA, 2016. ACM.

\bibitem{sundermeyer2012lstm}
M.~Sundermeyer, R.~Schl{\"u}ter, and H.~Ney.
\newblock Lstm neural networks for language modeling.
\newblock In {\em Thirteenth annual conference of the international speech
  communication association}, 2012.

\bibitem{tai2015improved}
K.~S. Tai, R.~Socher, and C.~D. Manning.
\newblock Improved semantic representations from tree-structured long
  short-term memory networks.
\newblock {\em arXiv preprint arXiv:1503.00075}, 2015.

\bibitem{VanAken:ottertune:sgmod:2017}
D.~Van~Aken, A.~Pavlo, G.~J. Gordon, and B.~Zhang.
\newblock Automatic database management system tuning through large-scale
  machine learning.
\newblock In {\em Proceedings of the 2017 ACM International Conference on
  Management of Data}, SIGMOD '17, pages 1009--1024, New York, NY, USA, 2017.
  ACM.

\bibitem{wu:qpp-concurrent-queries:vldb:2013}
W.~Wu, Y.~Chi, H.~Hac\'{\i}g\"{u}m\"{u}\c{s}, and J.~F. Naughton.
\newblock Towards predicting query execution time for concurrent and dynamic
  database workloads.
\newblock {\em Proc. VLDB Endow.}, 6(10):925--936, Aug. 2013.

\end{thebibliography}

\end{document}